\documentclass[apj]{emulateapj}
\usepackage{apjfonts}

\newcommand\HST{{\it{HST}}}
\newcommand\Chandra{{\it{Chandra}}}
\newcommand\phdn{\phantom{.0}}
\newcommand\phsn{\phantom{$-$0}}
\newcommand\phnn{\phantom{00}}
\newcommand\phsnn{\phantom{$-$00}}
\newcommand\ebv{\mbox{$E$($B\!-\!V$)}}
\newcommand\dg{\mbox{$^\circ$}}
\newcommand\kms{\mbox{km\ s$^{-1}$}}
\newcommand\ivol{\mbox{cm$^{-3}$}}
\newcommand\iarea{\mbox{cm$^{-2}$}}
\newcommand\muJybm{\mbox{$\mu$Jy/beam}}
\newcommand\Bn{\mbox{$B/n^{1/2}$}}
\newcommand\muGcm{\mbox{$\mu$G~cm$^{3/2}$}}
\newcommand\ergs{\mbox{erg~s$^{-1}$}}
\newcommand\ergscm{\mbox{erg~s$^{-1}$~cm$^{-2}$}}
\newcommand\ergscmA{\mbox{erg~s$^{-1}$~cm$^{-2}$~\AA$^{-1}$}}
\newcommand\xten[2]{\mbox{#1$\,\times\,10^{#2}$}}
\newcommand\aslit[2]{\mbox{#1\arcsec$\,\times\,$#2\arcsec}}
\newcommand\slit[2]{\mbox{#1$\,\times\,$#2}}
\def\la{\scalebox{1.2}[1.056]{\raisebox{1.2pt}{$\scriptscriptstyle\lesssim$}}}
\def\ga{\scalebox{1.2}[1.056]{\raisebox{1.2pt}{$\scriptscriptstyle\gtrsim$}}}
\makeatletter
\newcommand\wl{\mbox{$\lambda$}}
\newcommand\balmw[2]{\mbox{#1$\ $\wl#2}}
\newcommand\iona[2]{\mbox{#1$\,${\sc\@roman{#2}}}}
\newcommand\ionw[3]{\mbox{#1$\,${\sc\@roman{#2}}$\,$\wl#3}}
\newcommand\iondw[3]{\mbox{#1$\,${\sc\@roman{#2}}$\,$\wl\wl#3}}
\newcommand\forb[2]{\mbox{[#1$\,${\sc\@roman{#2}}]}}
\newcommand\forbw[3]{\mbox{[#1$\,${\sc\@roman{#2}}]$\,$\wl#3}}
\newcommand\forbdw[3]{\mbox{[#1$\,${\sc\@roman{#2}}]$\,$\wl\wl#3}}
\newcommand\forbww[4]{\mbox{[#1$\,${\sc\@roman{#2}}]$\,$\wl\wl#3,#4}}
\newcommand\semiforb[2]{\mbox{#1$\,${\sc\@roman{#2}}]}}
\newcommand\semiforbw[3]{\mbox{#1$\,${\sc\@roman{#2}}]$\,$\wl#3}}
\makeatother
\newcommand\NaD{\iona{Na}{1}$\,$D}
\newcommand\CaII{\iona{Ca}{2}}
\newcommand\HII{\iona{H}{2}}
\newcommand\Ha{\mbox{H$\alpha$}}
\newcommand\Haw{\balmw{\Ha}{6563}}
\newcommand\Hb{\mbox{H$\beta$}}
\newcommand\Hbw{\balmw{\Hb}{4861}}
\newcommand\Hg{\mbox{H$\gamma$}}
\newcommand\Hgw{\balmw{\Hg}{4340}}
\newcommand\Hd{\mbox{H$\delta$}}
\newcommand\Hdw{\balmw{\Hd}{4102}}
\newcommand\HeIIw{\ionw{He}{2}{4686}}
\newcommand\CIVw{\iondw{C}{4}{1549}}
\newcommand\NI{\forb{N}{1}}
\newcommand\NIw{\forbdw{N}{1}{5200}}
\newcommand\NII{\forb{N}{2}}
\newcommand\NIIwa{\forbw{N}{2}{6548}}
\newcommand\NIIwb{\forbw{N}{2}{6583}}
\newcommand\NIIww{\forbww{N}{2}{6548}{6583}}
\newcommand\OIwa{\forbw{O}{1}{6300}}
\newcommand\OIwb{\forbw{O}{1}{6364}}
\newcommand\OIww{\forbww{O}{1}{6300}{6364}}
\newcommand\OII{\forb{O}{2}}
\newcommand\OIIw{\forbdw{O}{2}{3727}}
\newcommand\OIII{\forb{O}{3}}
\newcommand\OIIIwa{\forbw{O}{3}{4959}}
\newcommand\OIIIwb{\forbw{O}{3}{5007}}
\newcommand\OIIIwc{\forbw{O}{3}{4363}}
\newcommand\OIIIww{\forbww{O}{3}{4959}{5007}}

\newcommand\NeIIIwa{\forbw{Ne}{3}{3869}}
\newcommand\NeIIIwb{\forbw{Ne}{3}{3968}}
\newcommand\NeIIIww{\forbww{Ne}{3}{3869}{3968}}
\newcommand\NeV{\forb{Ne}{5}}
\newcommand\NeVwa{\forbw{Ne}{5}{3346}}
\newcommand\NeVwb{\forbw{Ne}{5}{3426}}
\newcommand\SII{\forb{S}{2}}
\newcommand\SIIwa{\forbw{S}{2}{6716}}
\newcommand\SIIwb{\forbw{S}{2}{6731}}
\newcommand\SIIww{\forbww{S}{2}{6716}{6731}}
\newcommand\SIII{\forb{S}{3}}
\newcommand\SIIIwa{\forbw{S}{3}{9069}}
\newcommand\SIIIwb{\forbw{S}{3}{9532}}
\newcommand\ArIIIw{\forbw{Ar}{3}{7136}}

\journalinfo{To appear in {\it The Astrophysical Journal}, 2004 March 10}
\submitted{Received 2003 September 30; accepted 2003 November 24}
\shorttitle{NARROW-LINE REGION OF M51}
\shortauthors{BRADLEY, KAISER, \& BAAN}

\begin{document}

\title{Physical Conditions in the Narrow-Line Region of M51\footnotemark[1]}

\footnotetext[1]{Based on observations made with the NASA/ESA 
{\it Hubble Space Telescope}, obtained at the Space Telescope Science
Institute, which is operated by the Association of Universities for
Research in Astronomy under NASA contract NAS5-26555.}

\author{L. D. Bradley, M. E. Kaiser}
\affil{Department of Physics and Astronomy,
       The Johns Hopkins University, \\
       3400 N. Charles Street,
       Baltimore, MD 21218
}

\and

\author{W. A. Baan}
\affil{ASTRON, Westerbork Observatory,
       PO Box 2,
       Dwingeloo, 7990 AA,
       The Netherlands
}

\begin{abstract}

We have investigated the physical conditions in the narrow-line
region (NLR) of M51 using long-slit spectra obtained with the Space
Telescope Imaging Spectrograph (STIS) aboard the {\em Hubble Space
Telescope (HST)} and 3.6~cm radio continuum observations obtained
with the Very Large Array (VLA)\footnotemark[2].
Emission-line diagnostics were employed for
nine NLR clouds, which extend 2\farcs5 (102~pc) from the nucleus, to
examine the electron density, temperature, and ionization state of
the NLR gas.  The emission-line ratios are consistent with those
typically found in Seyfert nuclei and indicate that within the inner
near-nuclear region ($r\ \la\ 1\arcsec$) the ionization decreases
with increasing radius.  Upper-limits to the \OIII\ electron
temperature ($T_e\ \la\ 11,000$~K) for the inner NLR clouds indicate
that photoionization is the dominant ionization mechanism close to
the nucleus.  The emission-line fluxes for most of the NLR clouds can
be reproduced reasonably well by simple photoionization models using
a central power-law continuum source and supersolar nitrogen
abundances.  Shock+precursor models, however, provide a better fit to
the observed fluxes of an NLR cloud $\sim$2\farcs5 south of the
nucleus that is identified with the extra-nuclear cloud (XNC).  The
large \OIII\ electron temperature of this cloud ($T_e=24,000$~K)
further suggests the presence of shocks.  This cloud is straddled by
two radio knots and lies near the location where a weak radio jet,
$\sim$2\farcs5 (102~pc) in extent, connects the near-nuclear radio
emission with a diffuse lobe structure spanning
$\sim$4\arcsec\ (163~pc).  It is plausible that this cloud represents
the location where the radio jet impinges on the disk ISM.

\end{abstract}

\footnotetext[2]{The VLA is a facility of the National Radio
Astronomy Observatory, which is operated by Associated Universities,
Inc., under cooperative agreement with the National Science
Foundation.}

\addtocounter{footnote}{2}


\section{Introduction}

Photoionization by a central source is widely regarded as the
dominant process responsible for producing the emission-line spectra
observed in the Narrow-Line Region (NLR) of Seyfert galaxies.
Support for these photoionization models is provided by their general
success in reproducing the observed ultraviolet (UV) and optical
emission line spectra \citep[e.g.][]{Ferland1986} and by the observed
biconical morphology present in many Seyfert galaxies, which is
suggestive of collimation by a central ionizing source.

Many of these Seyfert galaxies also possess radio emission, which
appears collimated and extended, with similar orientation, on the
same scale as the optical emission.  The radio power has been shown
to be correlated with the \OIII\ luminosity and full-width
half-maximum (FWHM) \citep{deBruyn1978, Heckman1981, Whittle1985,
Whittle1992c}.  High spatial resolution imaging of the NLR often
illustrates a more detailed correspondence between the radio knots
and the optical emission-line clouds \citep{Pogge1995, Capetti1996,
Falcke1998}.

While photoionization models assuming a central ionizing source are
generally successful in reproducing the observed emission-line
spectra, there are some indications that other processes may also
contribute.  Simple photoionization models, in some cases,
underpredict the electron temperature in the gas (determined from the
\OIII\ ($\lambda4959\,+\,\lambda5007$)/$\lambda4363$ line ratio) and
are not always able to explain strong UV emission from intermediate
excitation line species, such as \semiforb{C}{3} and \iona{C}{4}
\citep{Ferland1986}.  Thus, the complex velocity distribution
presented by some Seyferts \citep{Cecil1988, Whittle1988, Winge1997,
Hutchings1998} and the aforementioned optical-radio correspondence
suggests that another process, such as shocks, may contribute to the
observed emission-line spectrum.  The role of shocks in the NLR
provides a key to the understanding the relative importance of
mechanical versus radiative energy processes in the active galactic
nuclei (AGN) model.

We examine the physical conditions influencing the NLR kinematics and
ionization structure, in particular the relative importance of
photoionization and shocks within the NLR of M51.  M51 (NGC~5194) is
a nearby \citep[8.4~Mpc,][]{Feldmeier1997}, nearly face-on, Sbc
grand-design spiral galaxy that hosts a weak active nucleus.  Early
observations attributed the ionization structure in the nuclear
region to photoionization by the active nucleus \citep{Rose1982,
Rose1983}.  However, later observations revealed structures that
indicated shocks also may be an important source of ionization in
this object \citep{Ford1985, Cecil1988}.  In particular, shocks were
postulated to be the dominant energy mechanism exciting the
emission-line gas in an extended extra-nuclear cloud (XNC) lying
$\sim$3\arcsec\ south of the nucleus.

Several factors led to this conclusion for the XNC.  \cite{Ford1985}
observed high electron temperatures determined from line-ratio
diagnostics, emission-line ratios qualitatively similar to that
observed in supernova remnants, a limb brightened morphology, and
abrupt changes in the spectral characteristics between the galactic
disk and the XNC cloud.  Further, \cite{Cecil1988} found
sharply-defined high-velocity emission structures that resembled the
termini of hydrodynamic flows and the broadest emission lines
localized along the northern boundary of the XNC.  A directed nuclear
outflow interacting with the disk of M51 was postulated as the source
of the shocks in the XNC.

Direct evidence for a weak radio jet connecting the nuclear radio
emission with the southern extra-nuclear emission was provided by
4.8~GHz observations obtained with the Very Large Array (VLA)
\citep{Crane1992}.  An extended lobe structure that is
morphologically similar to the optical XNC emission also is observed
in the radio map at this frequency.  A lower brightness halo engulfs
the nucleus, the jet, and regions north of the nucleus.

Recent {\em Chandra} observations ($\sim$1\arcsec\ resolution) by
\cite{Terashima2001} show an extended X-ray morphology for the XNC
that is similar to both the optical and radio data.  The X-ray data
further support the argument for shocks in the XNC.  These authors
modelled the X-ray spectra of this region with a thermal plasma
resulting from shock-heating due to outflows.  Their model yielded a
X-ray temperature $T_{X} = 6.7 \times 10^{6}$~K for the XNC, which
corresponds to a shock velocity of 690~\kms. This value is roughly
consistent with the $\sim$500~\kms\ shock velocity previously
inferred by \cite{Cecil1988}.

In this paper, we examine the physical conditions influencing the NLR
kinematics and ionization structure, in particular the relative
importance of photoionization and shocks within the NLR of M51.  In
\S~\ref{sec:obs} we present the observations, data reduction, and
stellar background subtraction.  In \S~\ref{sec:results} we present
the optical morphology, cloud kinematics, emission-line ratios, and
physical conditions within the NLR gas.  In \S~\ref{sec:radioobs} we
present our VLA 8.4~GHz radio observations and compare them with
archival {\em Hubble Space Telescope (HST)} Wide Field and Planetary
Camera (WFPC2) optical images.  We discuss the photoionization
modelling of the NLR clouds in \S~\ref{sec:photomodels}, and discuss
ionizing shock models in \S~\ref{sec:shocks}.  Our results are
summarized in \S~\ref{sec:summary}, and our conclusions are presented
in \S~\ref{sec:conclusions}.

Given the distance to M51 of 8.4~Mpc \citep{Feldmeier1997}, 1\arcsec\
corresponds to 40.7~pc.

\section{Observations}
\label{sec:obs}

\begin{figure*}[ht]
\epsscale{0.95}
\plotone{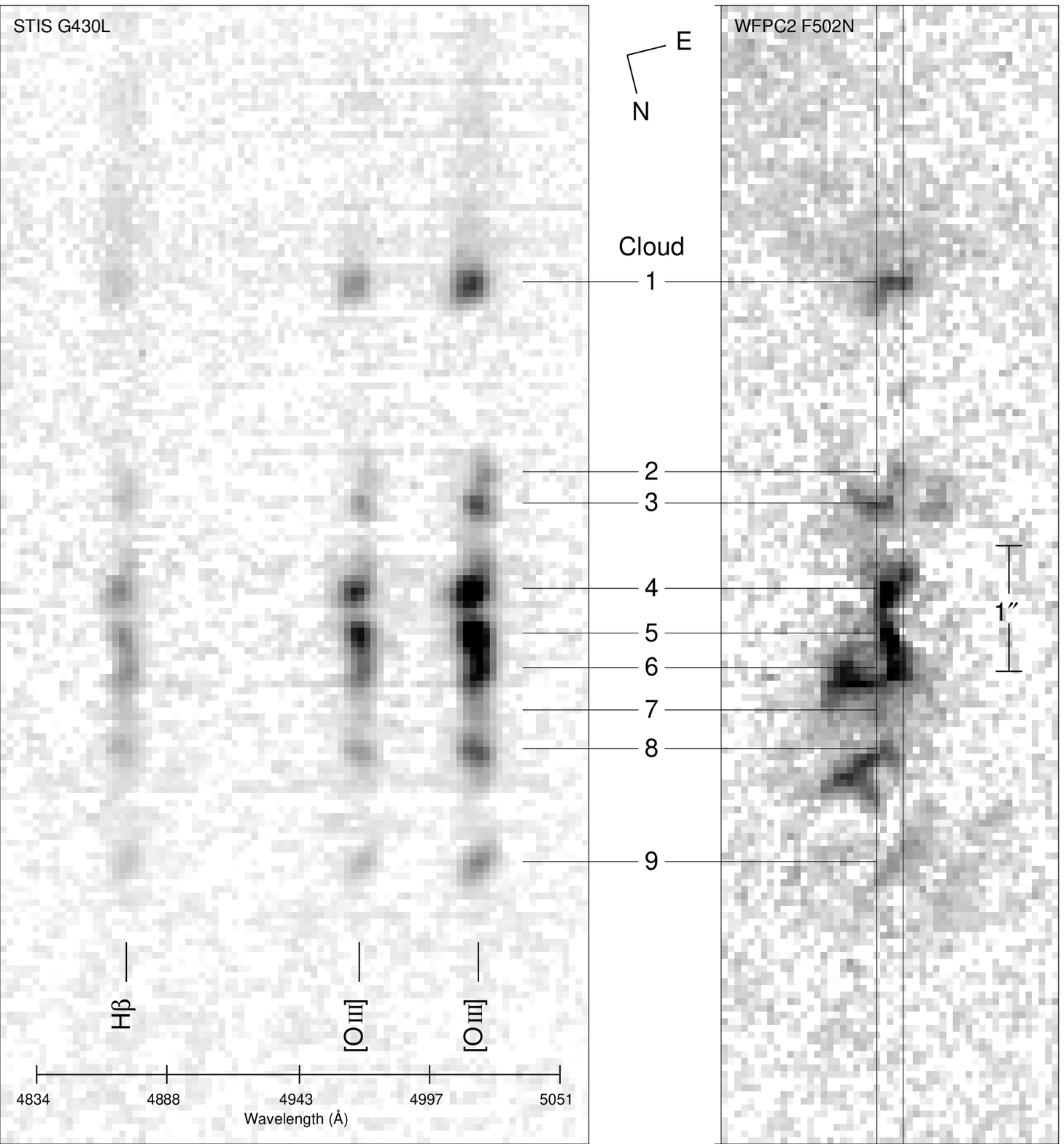}
\caption{(left) \Hb, \OIIIwa, and \OIIIwb\ emission lines extracted
from the background-subtracted \HST/STIS G430L spectral image (see
Table~\ref{tbl:obslog} for details) of the near-nuclear region of
M51.  The total spatial extent is 8\farcs6 (350~pc), and the
wavelength scale is in the observed frame.  (right)
Continuum-subtracted \HST/WFPC2 F502N narrowband \OIII\ image that
has been rotated to the STIS position angle (166\dg) and resampled to
the STIS plate scale (0\farcs051~pixel$^{-1}$).  The location of the
STIS \slit{52\arcsec}{0\farcs2} slit is overplotted on the WFPC2
image.  The NLR clouds have been labeled for identification in both
images.}
\label{fig:g430l}
\end{figure*}

\begin{figure*}[ht]
\epsscale{0.80}
\plotone{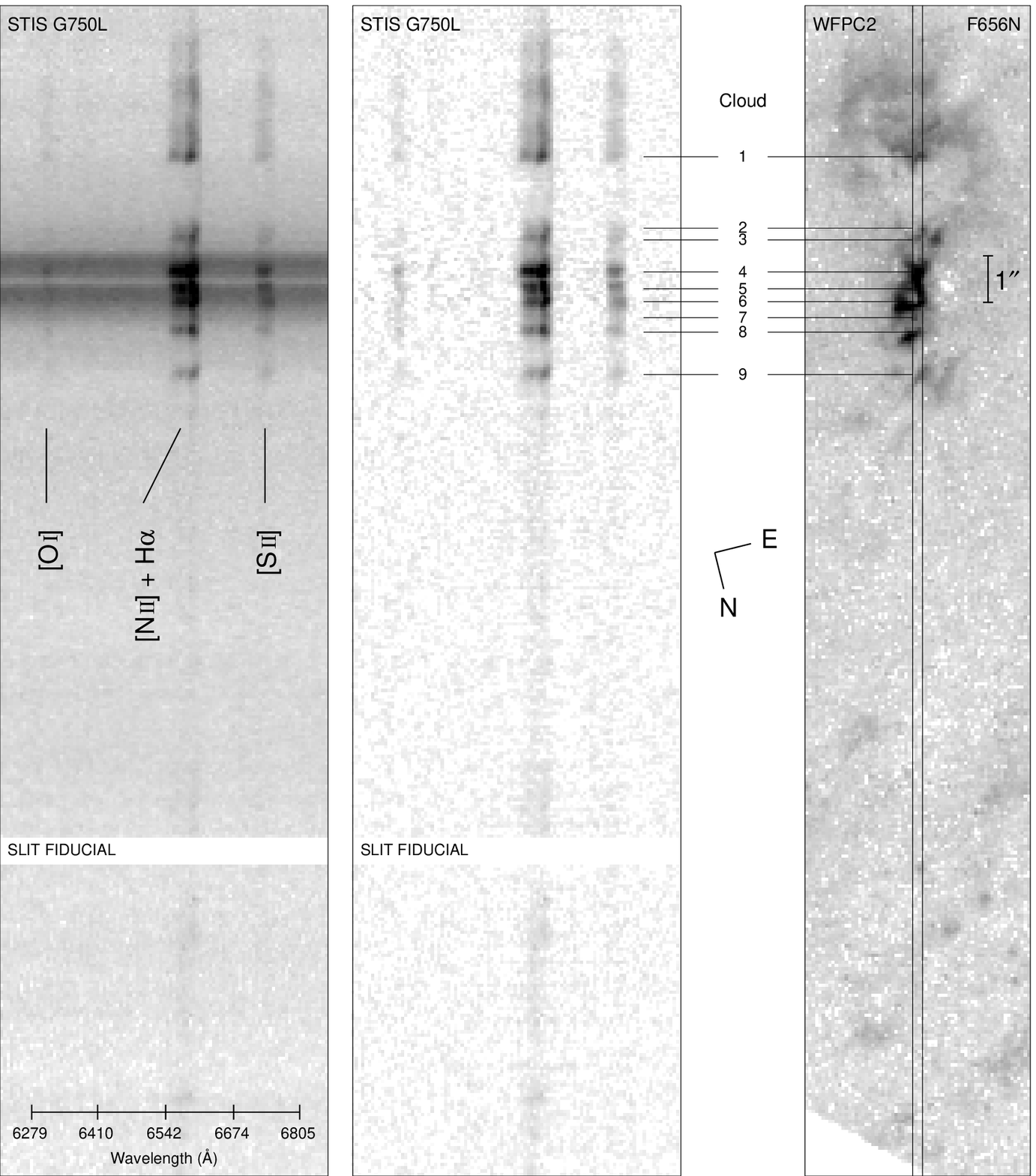}
\caption{(left) \OIww, \NIIww, \Ha, and \SIIww\ emission lines
extracted from the \HST/STIS G750L spectral image (see
Table~\ref{tbl:obslog} for details) of the near-nuclear region of
M51.  The total spatial extent is 23\farcs6 (960~pc), and the
wavelength scale is in the observed frame.  Note the extended
$\NII\,+\,\Ha$ emission to the north of the nucleus.  (middle)
Background-subtracted version of the spectral image at left.  (right)
Continuum-subtracted \HST/WFPC2 F656N narrowband $\NII\,+\,\Ha$ image
that has been rotated to the STIS position angle (166\dg) and
resampled to the STIS plate scale (0\farcs051~pixel$^{-1}$).  The
location of the STIS \slit{52\arcsec}{0\farcs2} slit is overplotted
on the WFPC2 image.  The NLR clouds have been labeled for
identification in both images.}
\label{fig:g750l}
\end{figure*}

High spatial resolution (0\farcs1, 4.1~pc) long-slit spectra of the
nuclear region of M51 were obtained with the Space Telescope Imaging
Spectrograph (STIS) aboard \HST.  The spectra were acquired through a
\slit{52\arcsec}{0\farcs2}\ slit oriented at a single position angle
of 166\dg\ (E of N).  This position angle was based on our analysis
of an archival \HST/WFPC2 F502N narrowband image of the near-nuclear
region of M51 and was selected to intersect the nucleus and several
of the brightest NLR clouds at this wavelength.  The low-dispersion
grating modes G430L (Fig.~\ref{fig:g430l}) and G750L
(Fig.~\ref{fig:g750l}) were used to provide continuous wavelength
coverage spanning the spectral region from 2900~\AA\ to 1~\micron.
The 0\farcs2 wide slit (0\farcs19 as-built width) projects onto
3.75~STIS CCD pixels (1.9 resolution elements) and corresponds to
spectral resolutions of 10.3~\AA\ for G430L and 18.3~\AA\ for G750L.
The associated velocity resolutions for this slit are 827~\kms\ at
\OIIw, 615~\kms\ at \OIIIwb, and 834~\kms\ at \Ha.

\begin{figure*}[ht]
\epsscale{1.0}
\plotone{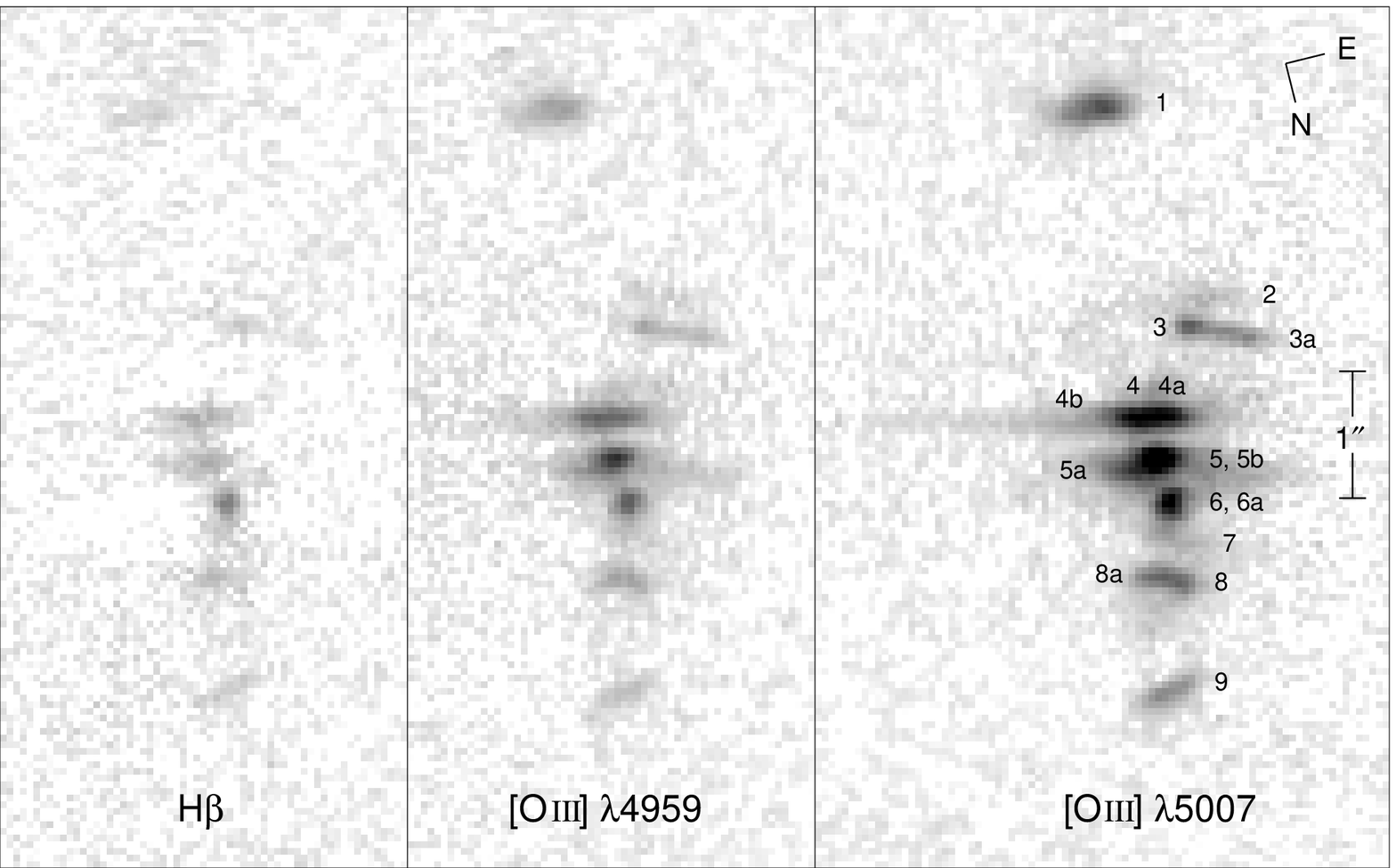}
\caption{Spectral cuts of the \Hb, \OIIIwa, and \OIIIwb\ emission
lines extracted from the background-subtracted \HST/STIS G430M
spectral image (see Table~\ref{tbl:obslog} for details).  The
clouds are labeled for identification.  Clouds with multiple
components have the lower-case letters ``a'' and ``b'' appended to
their cloud labels.}
\label{fig:g430m}
\end{figure*}

To determine the sub-arcsecond velocity structure of the
emission-line clouds, we observed the spectral region containing \Hb,
\OIIIwa, and \OIIIwb\ with the G430M ($\lambda_{\rm{c}}=4961$~\AA)
medium-dispersion grating mode (Fig.~\ref{fig:g430m}).  The spectral
resolution of this data is 1.0~\AA, which corresponds to 62~\kms\ at
\OIIIwb.

\begin{figure*}[p]
\epsscale{1.0}
\plottwo{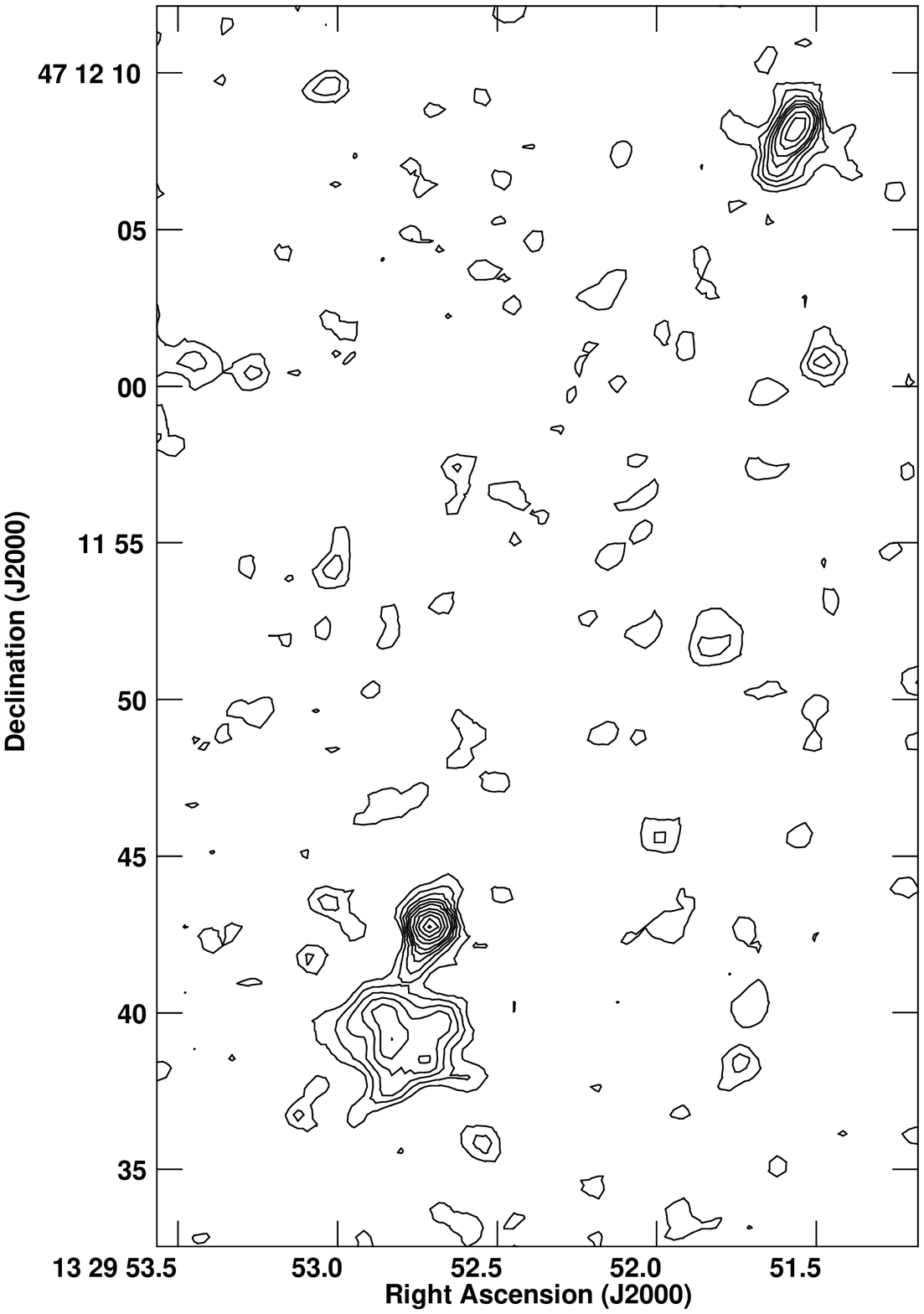}{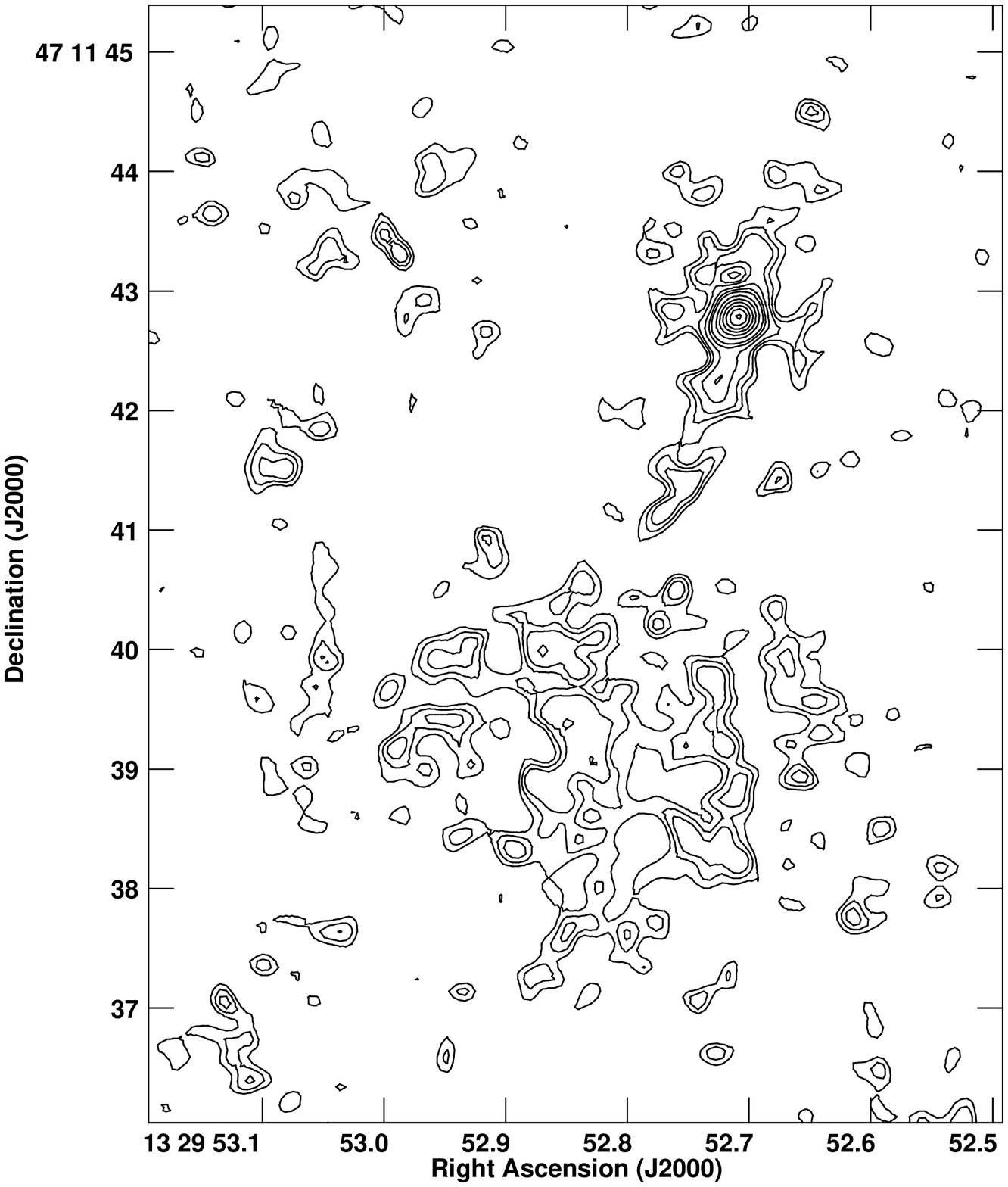}
\vspace{-7mm}
\plottwo{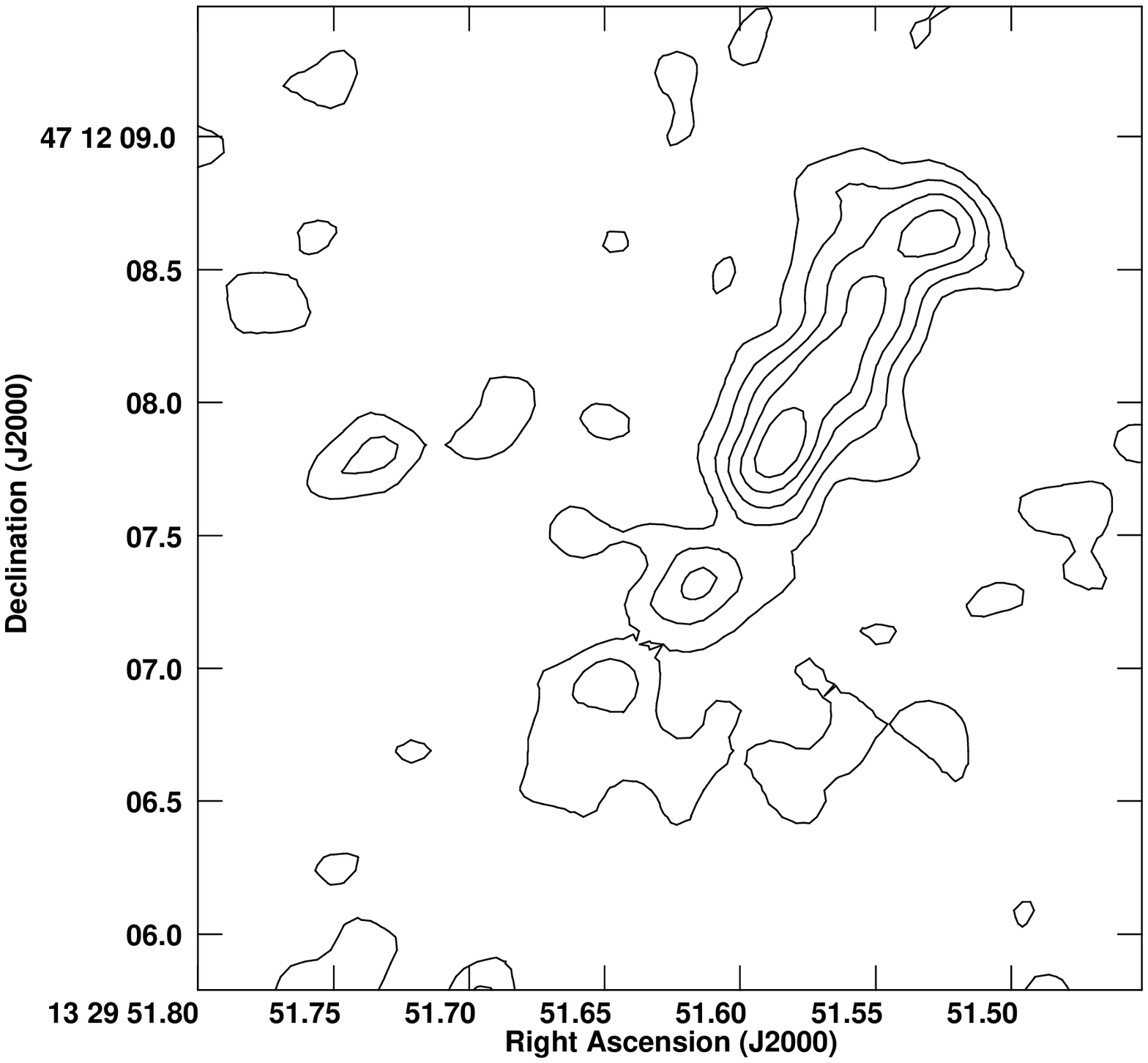}{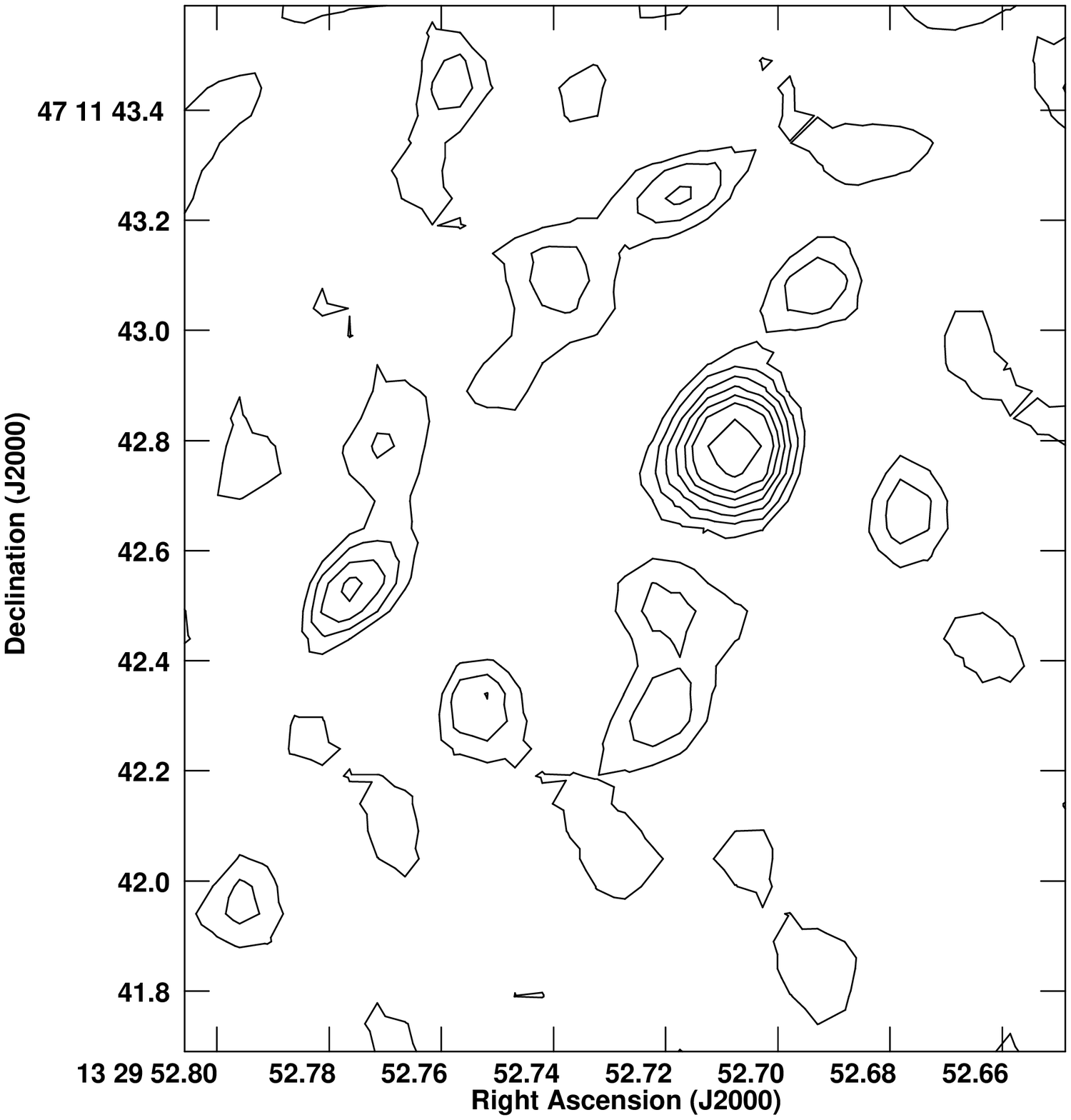}
\caption{\footnotesize
Radio structure in the nuclear region of M51 at 3.6~cm with the
VLA-A.  (a) (top left) The large scale nuclear structure in a
naturally weighted map at a resolution of \slit{0\farcs99}{0\farcs79}
and a rms noise of 13~$\mu$Jy/beam.  The contour levels are at
[2,4,6,8,10,12,16,20,24,28,32,36,40] $\times$ 10~$\mu$Jy/beam.  (b)
(top right) Naturally weighted map of the nuclear region with contour
levels at [2,4,6,8,10,12,16,20,24,28,32,36,40] $\times$
7.5~$\mu$Jy/beam.  The angular resolution is
\slit{0\farcs29}{0\farcs26} and the rms noise is 8~$\mu$Jy/beam.  (c)
(bottom left) Naturally weighted map of the northern source.  The
contour levels are the same as in (b).  The peak flux densities of
the sources, from north to south, are 64, 68, and 79~$\mu$Jy/beam.
(d) (bottom right) Uniformly weighted map of the nucleus at a
resolution of \slit{0\farcs18}{0\farcs15}.  The rms noise is
10~$\mu$Jy/beam and the contour levels are the same as in (a).}
\label{fig:radiomaps}
\end{figure*}

In addition, we have observed M51 with the VLA in A-array at 8.46~GHz
(3.6~cm) on 14 August 1999.  During the data reduction procedure in
AIPS, the data were thoroughly flagged in order to remove the effects
of a severe lightning storm that passed over the western arm during
the latter part of the observations.  Four representations of the
radio data of the central region of M51 are presented in
Figure~\ref{fig:radiomaps}.

A large scale map of the central region (Fig.~\ref{fig:radiomaps}a)
displays two prominent sources:  the nuclear source with an extended
lobe structure toward the south, and an elongated source at 27\farcs8
to the northwest at P.A.=335\dg.  The resolution of this naturally
weighted map is \slit{0\farcs99}{0\farcs79} and has a rms noise of
13~$\mu$Jy/beam.  Higher resolution maps are presented of both
sources.  The naturally weighted map of the nuclear source
(Fig.~\ref{fig:radiomaps}b) has a resolution of
\slit{0\farcs29}{0\farcs26} and a rms noise of 8~$\mu$Jy/beam.  The
uniformly weighted map of the nucleus (Fig.~\ref{fig:radiomaps}d) has
a resolution of \slit{0\farcs18}{0\farcs15} and a rms noise of
10~$\mu$Jy/beam.  While the naturally weighted map shows the nuclear
source together with the southern jet and the extended radio lobe,
the uniformly weighted map only shows the nuclear source and the
inner part of the southern jet.  The peak flux density of the nuclear
source in the two maps is 210~$\mu$Jy/beam and 195~$\mu$Jy/beam,
respectively.

The discrete northern source in the nuclear region of M51 consists of
three compact components and is presented in
Figure~\ref{fig:radiomaps}c.  The resolution of this naturally
weighted map is \slit{0\farcs29}{0\farcs26} and has a rms noise of
8.6~$\mu$Jy/beam.

A log of the optical and radio observations is presented in
Table~\ref{tbl:obslog}, which includes the archival \HST/WFPC2
\citep{Ford1996} and VLA 6~cm \citep{Crane1992} data.

\begin{deluxetable*}{lccccccc}
\tabletypesize{\footnotesize}
\tablecolumns{8}
\tablewidth{0pt}
\tablecaption{M51 Journal of Observations\label{tbl:obslog}}
\tablehead{\colhead{Date}
   &\colhead{Instrument} 
   &\colhead{Filter}
   &\colhead{Aperture} 
   &\colhead{Grating} 
   &\colhead{$\lambda_{c}$} 
   &\colhead{$\Delta\lambda$} 
   &\colhead{Exposure Time}\\
 & & & & & \multicolumn{1}{c}{(\AA\ or cm)} & \multicolumn{1}{c}{(\AA\ or cm)} & \multicolumn{1}{c}{(s or hr)} }
\startdata
\sidehead{Optical Data}
1995 Jan 24\tablenotemark{a} & WFPC2 & F502N &  &  & 5012 & \phn\phn26.9  & 1700 \\
1995 Jan 24\tablenotemark{a} & WFPC2 & F547M &  &  & 5446 & \phn486.6 & \phn860 \\
1995 Jan 24\tablenotemark{a} & WFPC2 & F656N &  &  & 6564 & \phn\phn21.5 & 1800 \\
1998 Apr 3  & STIS  & Clear & 50CCD           &       & 5852 & 4410\phdn & \phn\phn30   \\
1998 Apr 3  & STIS  & Clear & \slit{52\arcsec}{0\farcs2} &       & 5852 & 4410\phdn & \phn\phn30   \\
1998 Apr 3  & STIS  & Clear & \slit{52\arcsec}{0\farcs2} & G430L & 4300 & 2807\phdn & 3086 \\
1998 Apr 3  & STIS  & Clear & \slit{52\arcsec}{0\farcs2} & G430M & 4961 & \phn282\phdn & 2458 \\
1998 Apr 3  & STIS  & Clear & \slit{52\arcsec}{0\farcs2} & G750L & 7751 & 4987\phdn & 1644 \\
\sidehead{Radio Data}
1985 Jan 27\tablenotemark{b}    & VLA-A & & & & 6~cm   & 0.13~cm & 12 hr \\
1986 May  9-10\tablenotemark{b} & VLA-A & & & & 6~cm   & 0.13~cm & 12 hr \\
1986 May 11-12\tablenotemark{b} & VLA-A & & & & 6~cm   & 0.13~cm & 12 hr \\
1986 May 15-16\tablenotemark{b} & VLA-A & & & & 6~cm   & 0.13~cm & 12 hr \\
1999 Aug 14                     & VLA-A & & & & 3.6~cm & 0.04~cm & 10 hr \\[-1em]
\enddata 
\tablenotetext{a}{Archival data \citep{Ford1996}.}
\tablenotetext{b}{Archival data \citep{Crane1992}.}
\end{deluxetable*}

\subsection{Data Reduction}

The images and spectra were reduced using the IDL CALSTIS pipeline
developed for the STIS Instrument Development Team (IDT).  Cosmic
rays were removed from the data by combining all of the observations
in a given spectroscopic mode using an iterative sigma-clipping
algorithm based on a calibrated noise model of the images.  After
subtracting the detector bias and superdark frames, pixel-to-pixel
variations in the CCD response were removed from the G430L and G430M
spectra with the standard pipeline flat-field images.  A
contemporaneous flat-field image was used to remove the fringing in
the G750L spectral image at wavelengths longer than 7000~\AA.

The data were wavelength calibrated using exposures of an onboard
Pt/Cr-Ne spectral line lamp to determine the wavelength zero-point
offsets to the pipeline dispersion coefficients.  Geometric
rectification of the images provided a constant wavelength along each
column of the spectral images.  Using the pipeline sensitivity
tables, the data were flux-calibrated in units of \ergscmA\ per
cross-dispersion pixel.

Residual hot pixels present in featureless regions of the
post-pipeline processed spectral images were removed using an
iterative $\sigma$-clipping procedure.  A $1 \times 51$ pixel boxcar
was passed over these regions, and successive filtering iterations
were performed beginning with 8$\sigma$ and cleaning down to a
3$\sigma$ threshold.

The background present in the medium spectral resolution G430M image
consists of the near-nuclear stellar continuum of the M51 host
galaxy, which fills the entire length of the STIS slit.  Because
changes in the continuum shape are small ($\sim$5\%) over the narrow
G430M spectral range (282~\AA), we removed the background using an
empirical cross-dispersion template of the spectral image in regions
without emission features.  After masking the emission-line clouds,
the background template was generated by calculating the median of
each image row.  The cross-dispersion template was then subtracted
from each column of the spectral image.

Nine spatially resolved emission-line clouds were identified in the
data (see Figures~\ref{fig:g430l} and \ref{fig:g750l}).  Individual
cloud spectra were created by adding adjacent rows along the slit in
the spatial direction.  The number of summed rows (typically three
pixels, 0\farcs15) varied from cloud to cloud and was chosen based on
the spatial extent of the \OIIw\ emission line.

\subsection{G430L and G750L Host Galaxy Subtraction}

Because the active nucleus of M51 is obscured, which is expected for
a Seyfert~2 galaxy, the AGN continuum is not apparent in our lower
resolution G430L and G750L spectra.  However, the near-nuclear
spectra exhibit a strong stellar spectrum from the M51 host galaxy.
Since the \Hb, \Hg, and \Hd\ spectral features are comprised of both
emission and absorption lines, accurate measurement and subtraction
of the stellar components is important for accurately measuring the
emission-line flux.  This is also true for several other emission
lines (e.g. \NeIIIwa) that have nearby stellar absorption features.
\Ha\ absorption is not directly observed due to the strong adjacent
\NIIww\ emission lines.  However, the presence of \Hb\ stellar
absorption suggests that \Ha\ absorption also is present.
Measurement of the \Ha\ and \Hb\ stellar absorption features is
especially important because the \Ha/\Hb\ Balmer decrement is used as
our reddening indicator.  The removal of the stellar background is
complicated by the presence of variable extinction in the
near-nuclear region.  The variable reddening changes the continuum
shape of the spectra and adds an additional degeneracy to the well
known age-metallicity degeneracy.  If continuum shape alone was used
as a fit diagnostic, then a highly reddened spectrum could mimic an
older stellar population.  Thus, the stellar population background
must fit both the continuum shape and the stellar absorption lines.

To determine the stellar background in the near-nuclear region of
M51, we performed a population synthesis on our least extincted
(bluest) cloud spectrum in the inner nuclear region.  This cloud
(cloud 6) is 0\farcs43 from the obscured nucleus.  It is less
extincted than clouds 4 and 5, and exhibits very little differential
extinction across its spatial extent.  We did not expect a stellar
population gradient within the near-nuclear region ($\sim$1\arcsec,
41~pc) nor did we see evidence in the WFPC2 images for localized star
formation.  Consequently, cloud 6 was used to determine our stellar
population template in the near-nuclear region.  To increase the
signal-to-noise in the spectrum, we combined all four rows (0\farcs2)
of cloud 6 before performing the stellar population synthesis.
Assuming an invariant intrinsic stellar population within the
near-nuclear region, the only free parameters in fitting the
remaining emission-line clouds are the reddening, \ebv, and an
overall multiplicative scale factor.

The first step in the process was to separate the stellar and
emission-line contributions to the observed NLR spectrum.  In the
case where the emission line occurs in an otherwise featureless
spectral region, we excluded the emission line from the stellar
population fit.  In regions where the emission from the active
nucleus is superimposed upon stellar features, we simultaneously fit
the absorption and emission components using the SPECFIT procedure in
IRAF\footnote{IRAF is distributed by the National Optical Astronomy
Observatories, which is operated by the Association of Universities
for Research in Astronomy, Inc., under cooperative agreement with the
National Science Foundation.}.  These fits were performed on a
line-by-line basis for the spectral region encompassing the
absorption and emission lines.  A Gaussian profile was used for the
absorption, while the \OIIIwb\ emission-line profile was used as a
template for the emission.  The resulting absorption fit was inserted
into the spectral region of these two-component features.

The spectral synthesis fit was then performed on the ``stellar''
spectra using a reddened \citep{Fitzpatrick1999} linear combination
of empirical stellar templates \citep{Pickles1998}.  The
\cite{Pickles1998} stellar library was selected over the
\cite{Kurucz1993}, \cite{Silva1992}, and \cite{Bica1986} libraries
due to its complete wavelength coverage, 10~\AA\ resolution,
extensive stellar types, and the superiority of the resulting fit.
The 131 spectra in the Pickles stellar library spanned all luminosity
classes and included both metal-weak and metal-rich giants.  The
stellar synthesis represented the best linear combination of the
Pickles empirical stellar templates plus an overall reddening
correction.

To partially constrain the population synthesis, we used the results
of \cite{Panagia1996} on the dust and stellar content in the nuclear
region of M51.  \cite{Panagia1996} found that the stellar population
in the near-nuclear ``core'' of M51 ($r\ \la\ 1\arcsec$) has a
main-sequence turnoff around A0 (age $\ga$~400~Myr), while the
stellar population in the bulge ($r\ \ga\ 1\arcsec$) has a
main-sequence turnoff near G0 (age $\ga$~8~Gyrs).  Thus, for the
spectral synthesis on cloud 6 ($r = 0\farcs43$) we excluded O and B
main-sequence stars.

Before performing the stellar population fit, we normalized the NLR
cloud spectra at 5556~\AA\ to match the normalization of the stellar
templates.  It was unnecessary to broaden the absorption features in
the stellar templates since the stellar velocity dispersion in the
near-nuclear region of M51 \citep[62~\kms,][]{Delisle1992} is
negligible compared to our spectra resolution (615~\kms\ at \OIIIwb)
and that of the stellar templates (600~\kms\ at \OIIIwb).  The
\NaD\ absorption feature is contaminated by foreground Galactic
absorption \citep{Kormendy1996, Vaceli1997} and was excluded from the
stellar population fit.

\begin{figure}[ht]
\epsscale{1.2}
\plotone{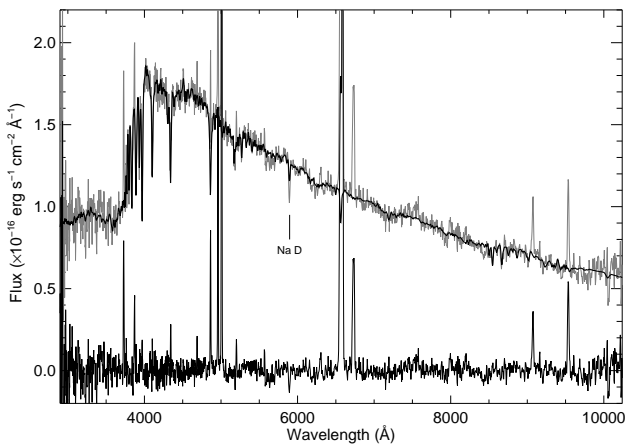}
\caption{Observed cloud 6 spectrum overplotted with the stellar
population fit (thick black line) used to subtract the underlying
stellar continuum.  The bottom spectrum shows the results of the
background subtraction.}
\label{fig:cloud6fit}
\end{figure}

The resulting best fit for cloud 6 (Fig.~\ref{fig:cloud6fit}) yields
a stellar \ebv\ of 0.09 and a $\chi^2_{\nu}=0.52$ ($\nu = 1506$).
The lower portion of Fig.~\ref{fig:cloud6fit} shows the cloud 6
spectrum after the stellar background was removed.  The
background-subtracted spectrum shows very little residual stellar
structure with the exception of \NaD.  The \NaD\ feature does not
impact our data because we do not observe any emission lines in this
region of the spectrum.

Comparing the stellar population fit to the near-nuclear NLR clouds
showed that the stellar composition appears constant within $r\
\la\ 1\arcsec$ (41~pc, clouds 3$-$7).  Therefore, we fit the cloud 6
composite stellar template to the NLR cloud spectra within this
region by varying the reddening of the stellar template and
multiplying it by an overall scale factor.

Because clouds 4 and 5 exhibit variable reddening across their
spatial extents (5 pixels, 0\farcs25), each of these clouds was split
into two cloudlets:  one that represented the co-addition of two
rows, and the other the co-addition of three rows.  The spectra were
binned in the spatial direction to both increase the signal-to-noise
and to perform the measurement on the scale of a spatial resolution
element (2 pixels, 0\farcs1).  All other emission-line clouds were
summed over their full spatial extents (typically three pixels,
0\farcs15) before fitting.

\begin{figure}[ht]
\epsscale{1.2}
\plotone{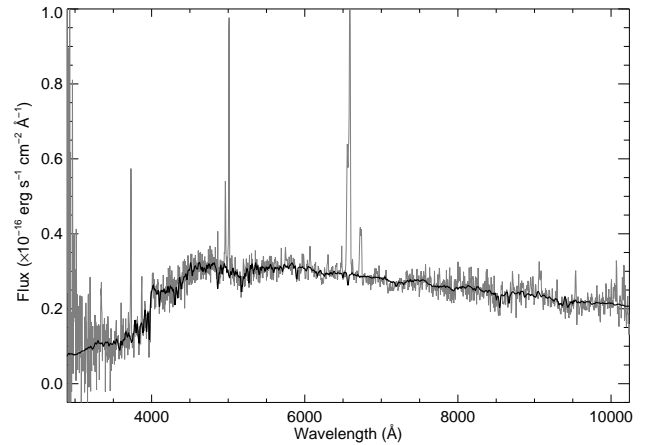}
\caption{Observed cloud 2 spectrum overplotted with the stellar
population fit (thick black line) used to subtract the underlying
stellar continuum.}
\label{fig:cloud2fit}
\end{figure}

While the stellar composition appears constant within the central
radius of $\sim$1\arcsec\ (41~pc), the near-nuclear stellar
population template provides a poorer fit to the NLR clouds at radii
$\ga$ 1\arcsec.  From inspection of the stellar background as a
function of radius, the relative strengths of the Balmer absorption
lines decrease while the \CaII\ and metal (e.g. TiO and MgH)
absorption lines increase, indicating an increased contribution from
an older stellar population outside the near-nuclear region.
Consequently, we performed a new stellar population synthesis for
clouds 2 and 8 ($r\ \ga\ 1\arcsec$).  Again using the results of
\cite{Panagia1996}, we excluded main-sequence stars earlier than G0
in the spectral synthesis of clouds 2 and 8.
Figure~\ref{fig:cloud2fit} shows the resulting stellar background fit
to cloud 2.

\begin{figure}[ht]
\epsscale{1.2}
\plotone{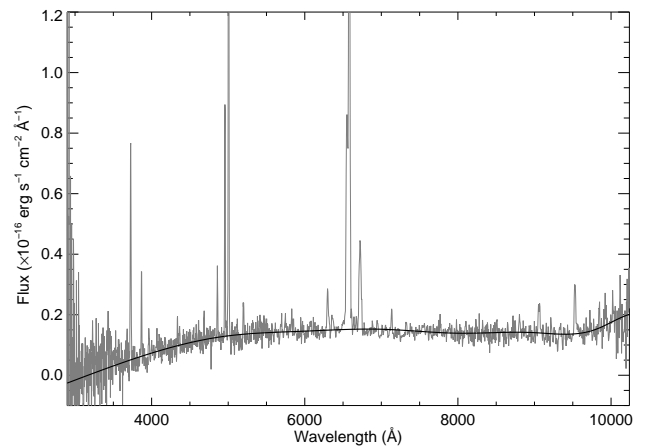}
\caption{Observed cloud 1 spectrum overplotted with the cubic
spline fit (thick black line) used to subtract the background
continuum.}
\label{fig:cloud1fit}
\end{figure}

Because we do not observe absorption features in the spectra of
clouds 1 and 9 ($r\ \ga\ 2\arcsec$), we fit the stellar background of
these clouds with a cubic spline.  The background fit to cloud 1 is
shown in Figure~\ref{fig:cloud1fit}.

\subsection{NLR Cloud Velocities}

Cloud velocities and velocity dispersions were measured using the
bright and unblended \OIIIwb\ line on the STIS G430M spectral image.
The central wavelength and FWHM of each \OIIIwb\ line was obtained by
fitting a single Gaussian plus a linear background to the spectral
region adjacent to the line.  Five clouds were not well fit by a
single Gaussian in the higher spectral resolution G430M data, most
likely due to a superposition of clouds along the line-of-sight.
These clouds were fit using multiple Gaussian components.  We label
the additional clouds by appending the lower-case letters ``a'' and
``b'' to the cloud identifications (see Fig.~\ref{fig:g430m}).

Several of the emission-line clouds were not centered within the STIS
slit.  We determined the relative position of the clouds within the
slit by using the \HST/WFPC2 F502N narrowband image of the
near-nuclear region of M51.  We rotated the WFPC2 F502N image to the
same position angle as our STIS observations (166\dg) and identified
the emission-line clouds present our STIS spectra.  Gaussian profiles
were then fit to the NLR clouds in the F502N image to determine the
difference between the cloud centroid and the center of the slit.
These pixel differences were converted to velocity shifts and were
used to correct the cloud radial velocities.  At a given position
along the slit multiple clouds are not spatially resolved in the
WFPC2 images.  Thus, the same slit offset correction was applied to
clouds with multiple components.

The intrinsic velocity dispersion of each cloud was calculated by
subtracting, in quadrature, the instrumental line broadening of
62~\kms\ from the measured FWHM velocity widths of the
\OIIIwb\ emission lines in the G430M spectral image.  We used the
\HST/WFPC2 F502N image to estimate the instrumental broadening for
clouds 2, 3, 5, and 6, which do not fully illuminate the slit width.
The instrumental broadening for these clouds was found to be
$\sim$45~\kms.

The G430M \OIIIwb\ cloud velocities and velocity dispersions are
presented in Table~\ref{tbl:cloudvel}.  The uncertainties were
propagated from the errors in the Gaussian fits and are given in
parenthesis.  A systemic velocity of
463~\kms\ \citep{deVaucouleurs1991} was subtracted from the radial
velocities to correct for the recessional motion of the M51 host
galaxy.

\begin{deluxetable}{ccc}
\tabletypesize{\normalsize}
\tablecolumns{3}
\tablewidth{0pt}
\tablecaption{Cloud Velocities and Velocity Dispersions\label{tbl:cloudvel}}
\tablehead{\colhead{Cloud}
   &\colhead{Radial Velocity}
   &\colhead{Velocity Dispersion}\\
   & \multicolumn{1}{c}{(\kms)} & \multicolumn{1}{c}{(\kms)}
}
\startdata
 1   &    $-$145 (1) &    107 \phn(3) \\
 2   &   \phsn85 (6) &    202    (18) \\
 3   &   \phs116 (1) & \phn25 \phn(9) \\
 3a  &   \phs217 (6) &    138    (13) \\
 4   & \phn$-$37 (7) & \phn66    (11) \\
 4a  &   \phsn41 (7) & \phn66    (10) \\
 4b  & \phn$-$31 (4) &    331    (13) \\
 5   &   \phsnn9 (1) & \phn37 \phn(2) \\
 5a  & \phn$-$15 (2) &    139 \phn(5) \\
 5b  &   \phsn45 (5) &    338    (11) \\
 6   &   \phsn30 (1) & \phn22 \phn(3) \\
 6a  &   \phsn24 (4) &    221    (13) \\
 7   &   \phsn67 (2) &    105 \phn(6) \\
 8   &   \phsn53 (7) & \phn69    (11) \\
 8a  & \phn$-$18 (9) & \phn49    (15) \\
 9   &   \phsn34 (1) & \phn74 \phn(3)
\enddata
\end{deluxetable}

\subsection{Emission-Line Flux Measurements}

The emission-line fluxes were measured using the IRAF routine
SPECFIT.  Isolated emission lines were measured using a Gaussian
profile with a linear background.  Since we detected weak residual,
broad TiO absorption in the spectral region near the \OIww\ lines, a
quadratic background was fit to the local continuum near these lines
and subtracted prior to fitting the emission.  The \ArIIIw\ flux was
measured by simple numerical integration across the line profile due
to the weakness of this emission line and uncertainties in the TiO
stellar absorption.  The G430L \OIIIwb\ profile was used as a
template to deblend the $\NII\,+\,\Ha$ lines and the \SIIww\ doublet
in the G750L spectral images.  Because the G430L and G750L grating
modes have different spectral resolutions, the width of the G430L
\OIII\ profile was allowed to vary in the fits.  Using the
\OIII\ profile to fit the \Ha\ line allowed for one less free
parameter and resulted in more reliable fits to the blended
$\NII\,+\,\Ha$ complex.  For cases in which both the Gaussian and
\OIII\ template fit gave reliable results, the \Ha\ fluxes agreed
within 2\%.  The \NIIww\ doublet lines were constrained to have the
same FWHM velocity widths, fixed relative central wavelengths, and an
intensity ratio fixed by atomic physics (1:2.95).  The \Ha\ FWHM was
fixed to the value of the \NIIww\ lines.  Similarly, the
\SIIww\ doublet lines were constrained to have fixed relative central
wavelengths and identical FWHM velocity widths.

\subsection{Reddening Correction}

We used the observed Balmer decrement of the \Ha/\Hb\ line ratio to
estimate the reddening of each NLR cloud.  The reddening curve of
\cite{Fitzpatrick1999} was applied assuming an intrinsic \Ha/\Hb\
ratio of 3.1.  While case B recombination yields \Ha/\Hb=2.85 for
$n_{e}=10^{4}$~\ivol\ and $T_{e}=10^{4}$~K \citep{Brocklehurst1971},
an assumed ratio of \Ha/\Hb=3.1 is more appropriate due to
collisional excitation of \Ha\ in AGN environments \citep[and
references therein]{Ho1993}.  The largest values of \ebv\
($\sim$0.37, $A_V=1.1$~mag) are found for the two innermost NLR
clouds (4 and 5), which straddle the central dust lane.

To account for the differential reddening across clouds 4 and 5, each
cloudlet spectrum was separately dereddened before summing the cloud
flux.  The values of \ebv\ and the dereddened emission-line fluxes,
normalized to \Hb, are presented in Table~\ref{tbl:deredflux}.  The
propagated dereddened flux errors are given in parentheses.

\begin{turnpage}
\begin{deluxetable*}{lccccccccc}
\tabletypesize{\footnotesize}
\tablecolumns{10}
\tablewidth{0pt}
\tablecaption{Dereddened Emission-Line Fluxes (normalized to \Hb)\label{tbl:deredflux}}
\tablehead{\colhead{Line}
   &\colhead{Cloud 1}
   &\colhead{Cloud 2}
   &\colhead{Cloud 3}
   &\colhead{Cloud 4}
   &\colhead{Cloud 5}
   &\colhead{Cloud 6}
   &\colhead{Cloud 7}
   &\colhead{Cloud 8}
   &\colhead{Cloud 9}
}
\startdata
\NeVwa &   \nodata  &   \nodata  &   \nodata  & \phn0.18 (0.04) & \phn0.23 (0.05) &   \nodata  &   \nodata  &   \nodata  &   \nodata  \\
\NeVwb &   \nodata  &   \nodata  &   \nodata  & \phn0.40 (0.10) & \phn0.98 (0.14) &   \nodata  &   \nodata  &   \nodata  &   \nodata  \\
\OIIw & \phn4.39 (2.85) & \phn5.57 (4.27) & \phn3.24 (1.65) & \phn1.91 (0.33) & \phn1.23 (0.15) & \phn1.26 (0.28) & \phn1.93 (0.96) & \phn2.63 (1.16) & \phn1.92 (1.29) \\
\NeIIIwa & \phn1.46 (0.93) & \phn1.35 (1.04) & \phn0.83 (0.48) & \phn0.99 (0.17) & \phn1.10 (0.14) & \phn0.62 (0.15) &   \nodata  & \phn0.97 (0.43) & \phn0.43 (0.33) \\
\NeIIIwb & \phn0.44 (0.27) & \phn0.40 (0.29) &   \nodata  & \phn0.29 (0.05) & \phn0.32 (0.04) & \phn0.18 (0.04) &   \nodata  & \phn0.29 (0.12) &   \nodata  \\
\Hdw & \phn0.18 (0.13) &   \nodata  &   \nodata  & \phn0.23 (0.06) & \phn0.31 (0.05) & \phn0.26 (0.07) &   \nodata  &   \nodata  &   \nodata  \\
\Hgw & \phn0.45 (0.28) & \phn0.70 (0.54) &   \nodata  & \phn0.59 (0.10) & \phn0.43 (0.06) & \phn0.42 (0.10) & \phn0.49 (0.26) & \phn0.27 (0.14) &   \nodata  \\
\OIIIwc & \phn0.50 (0.32) &   \nodata  &   \nodata  &   \nodata  &   \nodata  &   \nodata  &   \nodata  &   \nodata  &   \nodata  \\
\HeIIw & \phn0.41 (0.25) &   \nodata  & \phn0.33 (0.19) & \phn0.25 (0.04) & \phn0.32 (0.04) & \phn0.23 (0.06) & \phn0.34 (0.17) & \phn0.34 (0.14) & \phn0.32 (0.21) \\
\Hbw & \phn1.00 (0.57) & \phn1.00 (0.68) & \phn1.00 (0.45) & \phn1.00 (0.14) & \phn1.00 (0.10) & \phn1.00 (0.19) & \phn1.00 (0.43) & \phn1.00 (0.39) & \phn1.00 (0.58) \\
\OIIIwa & \phn3.92 (2.18) & \phn1.74 (1.14) & \phn2.44 (1.05) & \phn2.77 (0.38) & \phn3.80 (0.36) & \phn1.78 (0.33) & \phn1.02 (0.43) & \phn2.13 (0.80) & \phn1.77 (1.00) \\
\OIIIwb & 11.27 (6.23) & \phn4.96 (3.24) & \phn6.92 (2.97) & \phn7.87 (1.08) & 10.76 (1.02) & \phn5.11 (0.94) & \phn2.94 (1.23) & \phn6.11 (2.29) & \phn5.10 (2.87) \\
\NIw & \phn0.80 (0.44) & \phn0.48 (0.32) & \phn0.32 (0.15) & \phn0.43 (0.06) & \phn0.26 (0.03) & \phn0.31 (0.07) & \phn0.44 (0.20) & \phn0.58 (0.22) & \phn0.25 (0.15) \\
\OIwa & \phn1.49 (0.73) & \phn0.66 (0.40) & \phn0.52 (0.21) & \phn0.59 (0.08) & \phn0.35 (0.04) & \phn0.43 (0.09) & \phn0.64 (0.27) & \phn0.61 (0.21) & \phn1.10 (0.56) \\
\OIwb & \phn0.48 (0.23) &   \nodata  &   \nodata  & \phn0.18 (0.02) & \phn0.11 (0.01) & \phn0.14 (0.02) &   \nodata  & \phn0.19 (0.06) &   \nodata  \\
\NIIwa & \phn4.79 (2.29) & \phn2.89 (1.64) & \phn2.18 (0.81) & \phn3.15 (0.37) & \phn2.20 (0.18) & \phn2.32 (0.37) & \phn2.53 (0.92) & \phn3.10 (1.01) & \phn2.82 (1.37) \\
\Haw & \phn3.09 (1.51) & \phn3.07 (1.76) & \phn3.06 (1.15) & \phn3.06 (0.37) & \phn3.06 (0.25) & \phn3.09 (0.50) & \phn2.88 (1.05) & \phn3.09 (1.02) & \phn3.09 (1.54) \\
\NIIwb & 14.33 (6.85) & \phn8.30 (4.70) & \phn6.32 (2.36) & \phn9.27 (1.10) & \phn6.53 (0.54) & \phn6.88 (1.11) & \phn7.55 (2.74) & \phn9.09 (2.96) & \phn8.21 (4.02) \\
\SIIwa & \phn1.90 (0.92) & \phn1.19 (0.68) & \phn0.88 (0.33) & \phn0.99 (0.12) & \phn0.88 (0.08) & \phn1.06 (0.19) & \phn1.08 (0.43) & \phn1.13 (0.39) & \phn1.31 (0.70) \\
\SIIwb & \phn1.71 (0.84) & \phn1.14 (0.65) & \phn0.90 (0.34) & \phn1.06 (0.13) & \phn0.88 (0.08) & \phn1.15 (0.21) & \phn1.26 (0.48) & \phn1.34 (0.44) & \phn1.34 (0.68) \\
\ArIIIw & \phn0.49 (0.24) &   \nodata  & \phn0.23 (0.10) & \phn0.19 (0.03) & \phn0.25 (0.02) & \phn0.22 (0.06) &   \nodata  & \phn0.43 (0.15) & \phn0.32 (0.19) \\
\SIIIwa & \phn0.96 (0.44) &   \nodata  &   \nodata  & \phn0.54 (0.06) & \phn0.65 (0.06) & \phn0.72 (0.12) & \phn0.68 (0.31) & \phn0.63 (0.21) & \phn0.53 (0.25) \\
\SIIIwb & \phn1.47 (0.64) & \phn0.95 (0.51) & \phn0.88 (0.30) & \phn1.14 (0.12) & \phn1.22 (0.09) & \phn1.26 (0.19) & \phn1.29 (0.47) & \phn1.41 (0.42) & \phn0.73 (0.34) \\[2ex]
F(\Hb)\tablenotemark{a} & \phn3.16 (1.27) & \phn3.54 (1.70) & \phn7.66 (2.43) & 36.83 (3.70) & 35.73 (2.46) & 11.62 (1.57) & \phn2.32 (0.71) & \phn4.43 (1.21) & \phn2.03 (0.83) \\
\ebv & \phn0.09 (0.12) & \phn0.32 (0.14) & \phn0.40 (0.09) & \phn0.35 (0.03) & \phn0.38 (0.02) & \phn0.11 (0.04) & \phn0.00 (0.09) & \phn0.15 (0.08) & \phn0.10 (0.12) \\[-1em]
\enddata
\tablecomments{Fluxes are relative to \Hb.  Errors are given in parenthesis.}
\tablenotetext{a}{Dereddened \Hb\ Flux ($\times 10^{-16}$ \ergscm).}
\end{deluxetable*}
\end{turnpage}

\section{Results}
\label{sec:results}

\subsection{NLR Morphology \label{sec:ioncone}}

\begin{figure*}[th]
\epsscale{1.0}
\plottwo{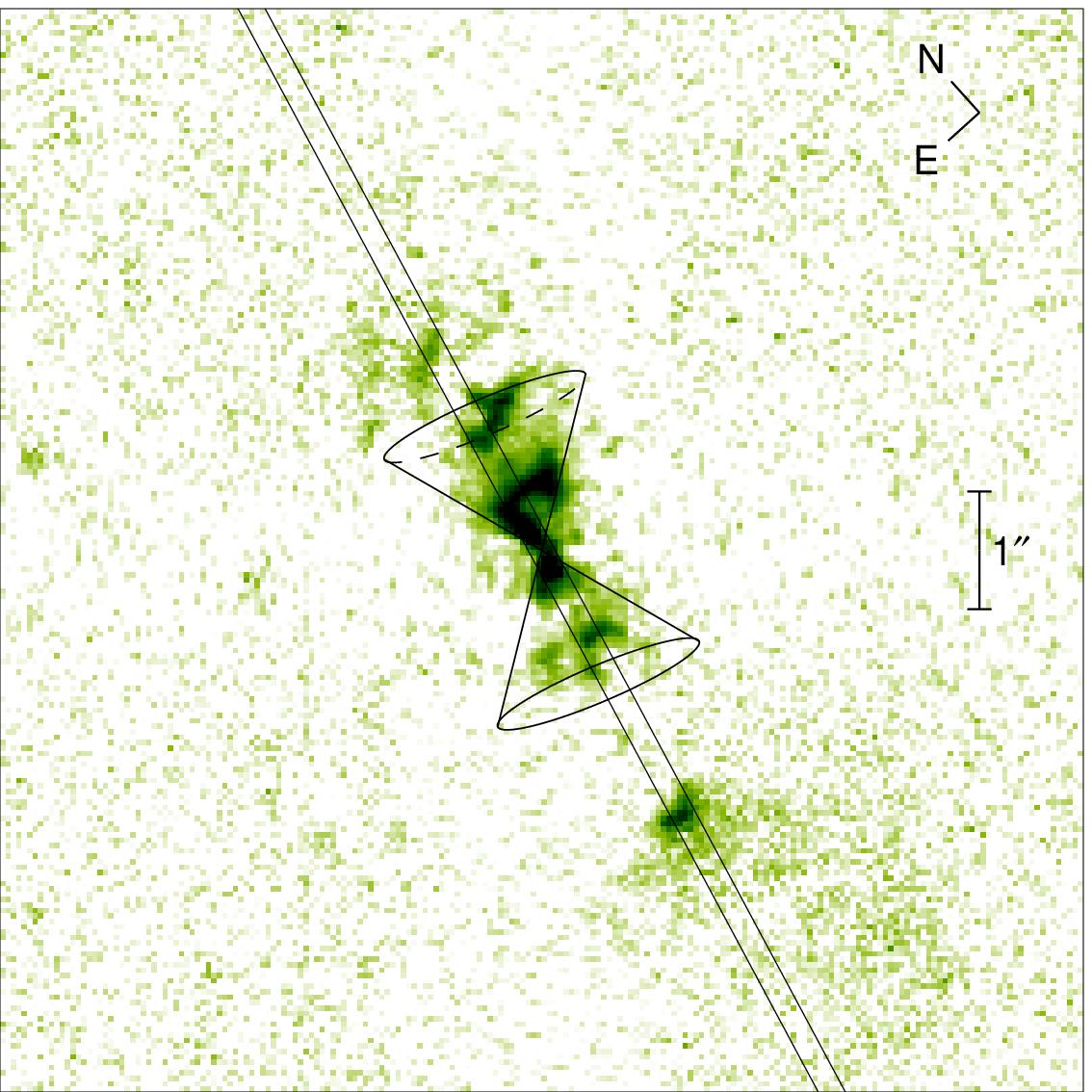}{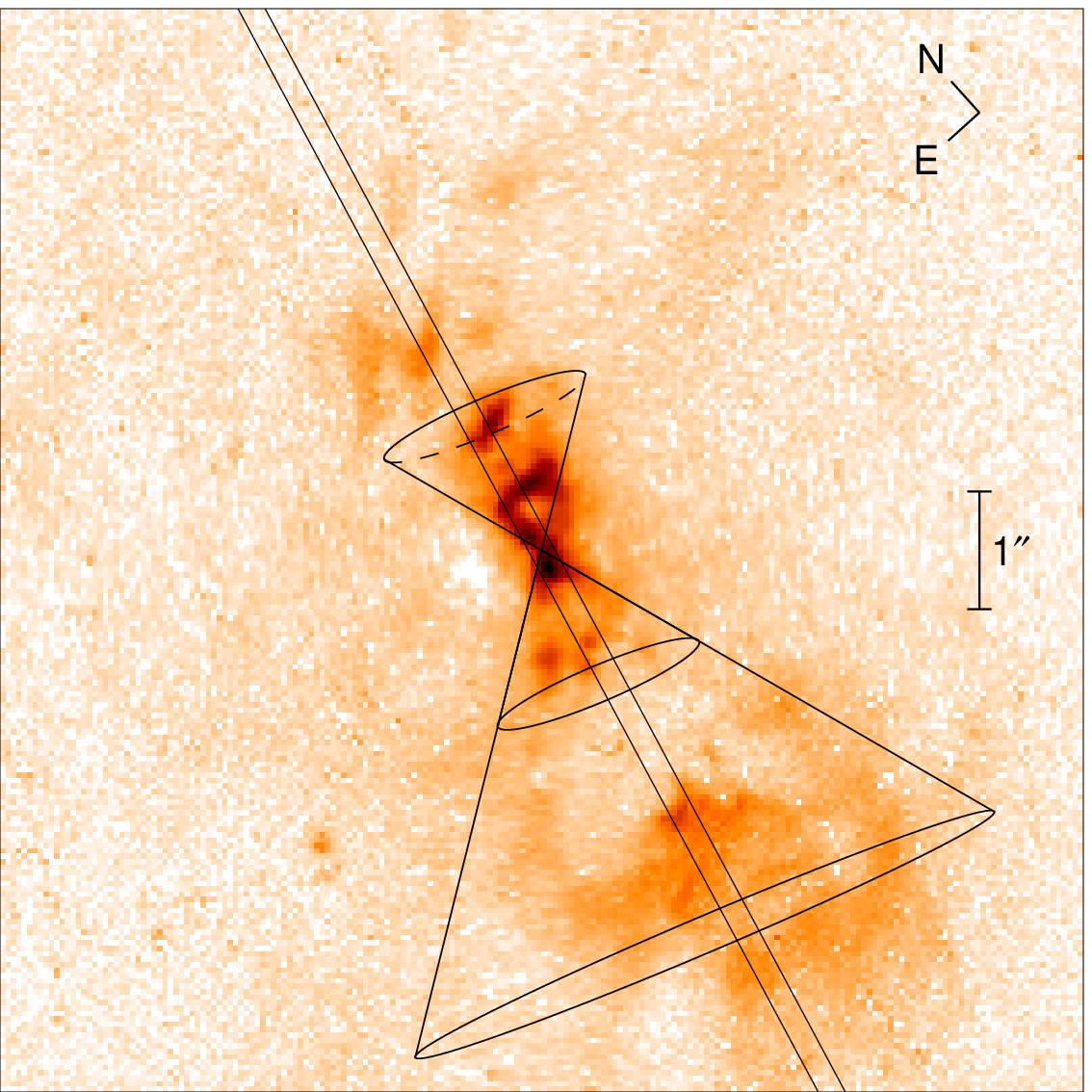}
\caption{(a) Continuum-subtracted \HST/WFPC2 F502N narrowband
\OIII\ image and (b) \HST/WFPC2 F656N narrowband $\NII\,+\,\Ha$
image showing the biconical emission.  The overplotted bicone has
a projected opening angle of 74\dg\ with an axis position angle of
163\dg.  The location of the \slit{52\arcsec}{0\farcs2} slit for
our STIS observations (P.A.=166\dg) also is shown.}
\label{fig:bicone}
\end{figure*}

The continuum-subtracted \HST/WFPC2 narrowband \OIII\
(Fig.~\ref{fig:bicone}a) and $\NII\,+\,\Ha$ images
(Fig.~\ref{fig:bicone}b) of the near-nuclear region of M51 suggest
the presence of a biconical morphology for the inner NLR gas.  In the
AGN paradigm, ionization cones result from the collimation of the
nuclear continuum radiation field by a dense, optically-thick
molecular torus and have been previously observed in several Seyfert
galaxies \citep{Pogge1988, Evans1993, Wilson1994}.  Most of the
emission in the near-nuclear region of M51 appears confined to an
ionization cone with a projected opening angle of $\sim$74\dg\ with
an axis position angle of $\sim$163\dg\ (see
Figures~\ref{fig:bicone}a and \ref{fig:bicone}b).

On the southern side of the nucleus, the $\NII\,+\,\Ha$ image
(Fig.~\ref{fig:bicone}b) shows a shell-like structure whose boundary
agrees well with the boundaries defined by the inner ionization
cone.  The eastern shell boundary is more prominent and better
defined than the western side, which appears to be fragmented.
Inside of the shell-like structure, we find a hollow cavity that is
occupied by the southern radio jet.  The most probable origin of the
cavity is that it has been evacuated by the radio jet (see
\S~\ref{sec:radioobs}).  Thus, the edges of the southern cone
structure may be dominated by shocks expanding laterally from the
radio jet instead of photoionization from the nuclear continuum
source.  While our single slit position does not intersect these
boundaries, \HST\ GO program 9147 (Ferruit et al., in preparation)
will map the emission near the radio jet to determine the source of
ionization in these structures.

Additionally, the overall position angle of the optical emission is
closely aligned with the observed radio emission at 3.6~cm and 6~cm
\citep{Crane1992}.  Recent CO imaging of M51 \citep{Scoville1998} has
revealed dense molecular gas confined to a disk that is associated
with the more opaque arm of the \mbox{\textsf{X}-shaped} dust
absorption feature that crosses the nucleus.  This molecular gas is
nearly perpendicular to the radio jet \citep{Ford1996, Grillmair1997}
and has a density $\ge\,10^{5}$~\ivol, which is sufficient to
collimate both the ionizing continuum and the radio-emitting plasma
\citep{Scoville1998}.  This putative molecular torus may also provide
a reservoir of matter to feed the active nucleus.

Extensive kinematic modelling by \cite{Cecil1988} indicates that the
jet axis is oriented $\sim$70\dg\ with our line-of-sight.  If the
radio jet and optical bicone axis are similarly collimated and nearly
co-spatial, then the resulting inclination angle for the ionization
cone is $\sim$20\dg, with the southern cone being nearer.

\subsection{Kinematics}

\begin{figure}[ht]
\epsscale{1.2}
\plotone{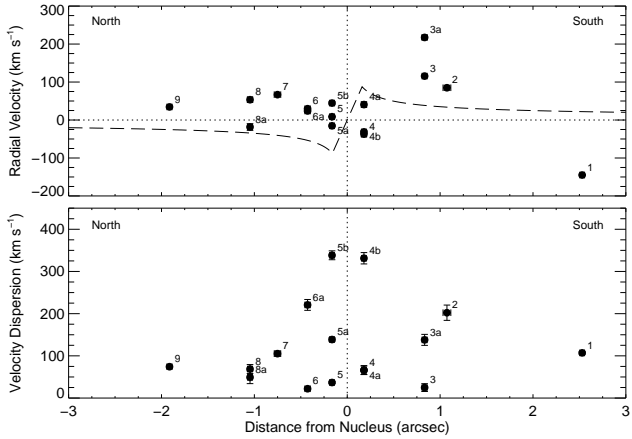}
\caption{Radial velocities (upper panel) and velocity dispersions
(lower panel) of the NLR clouds as a function of projected radial
distance from the nucleus.  The data points are labeled with the
cloud identifications.  The dashed line represents a simple
inclined Keplerian disk model in which the clouds are assumed to
lie in the plane of the galaxy ($i=20\dg$) and are undergoing
gravitational rotation about a central mass of $1 \times
10^8\ M_{\odot}$.  This model clearly does not fit the observed
NLR kinematics.}
\label{fig:cloudvel}
\end{figure}

We plot the cloud radial velocities and velocity dispersions as a
function of projected radial distance from the nucleus in
Figure~\ref{fig:cloudvel}.  The amplitude of the NLR cloud radial
velocities are relatively small and lie in the range from
$-$145~\kms\ (cloud 1) to 217~\kms\ (cloud 3a).  We find the smallest
radial velocities ($|v|\ \le\ 67$~\kms) for the nine clouds north of
the nucleus (clouds 5$-$9).  Only five of the NLR clouds (1, 4, 4b,
5a, and 8a) have blueshifted radial velocities, while the remaining
11 clouds are redshifted.  The highest velocity dispersions ($\Delta
v \approx 340~\kms$) are found closest to the nucleus (clouds 4b and
5b), and the upper-envelope of velocity dispersions decreases as the
projected radial distance increases.

In Figure~\ref{fig:cloudvel} we have overplotted a simple inclined
Keplerian disk model by assuming the NLR clouds lie in the plane of
the M51 host galaxy \citep[inclination=20\dg, eastern side nearer,
with major axis along P.A.=170\dg,][]{Tully1974b} and are undergoing
gravitational rotation about a central mass of $1 \times
10^8\ M_{\odot}$.  It is evident that both qualitatively and
quantitatively this model does not fit the observed radial velocity
field of the NLR clouds.  Most of the clouds north of the nucleus are
redshifted, however, we would expect them to exhibit blueshifted
radial velocities if they were located within the galactic disk.  The
lack of any obvious \textsf{S}-shape in the radial velocity
distribution suggests that the NLR clouds are not undergoing
gravitational rotation in a circumnuclear disk.

An alternative explanation for the kinematic structure is that the
NLR clouds are outflowing (or inflowing) from the nucleus.  Radial
outflow motions have been observed in several studies of the NLR in
Seyfert galaxies \citep[e.g.][]{Kaiser2000, Crenshaw2000, Ruiz2001}.
The relatively small observed radial velocities on the northern side
of the nucleus are consistent with an outflow (or inflow) that lies
nearly in the plane of the sky.  This picture is consistent with the
clouds lying near the front half of the ionization cone (see
Figures~\ref{fig:bicone}a and \ref{fig:bicone}b), which we have also
inferred to lie nearly in the plane of the sky.  While our kinematic
data cannot conclusively discriminate between outflow and inflow
radial motions, the presence of the southern radio jet and a possible
northern counter-jet favor the outflow scenario.

Clouds 1, 3, and 3a, located on the southern side of the nucleus,
possess the largest observed radial velocities ($|v|\ \ge\
116$~\kms).  Interestingly, these clouds lie on the same side of the
nucleus as the radio jet and have nearly coincident radio knots (see
\S~\ref{sec:radioobs}).  While cloud 1 is consistent with entrainment
by the jet outflow, clouds 2, 3, and 3a are not.  For a jet
orientation of 70\dg\ with the line-of-sight \citep{Cecil1988}, these
clouds should exhibit blueshifted radial velocities if they are
entrained by the jet.  A possible explanation for the larger radial
velocities of these clouds is that they lie behind the radio jet and
have offset velocities due to a laterally expanding flow away from
the radio jet.  If there was a weak northern counter jet, which has
since disrupted, we would expect clouds north of the nucleus to be
redshifted and those south of the nucleus to be blueshifted.  We see
a very weak signature supporting this scenario in both our kinematic
and radio observations.

Assuming that cloud 1 is entrained by the jet outflow (70\dg\ to the
line-of-sight), the resulting deprojected velocity of this cloud is
424~\kms.  This velocity is in agreement with the previously inferred
shock velocity of $\sim$500~\kms\ \citep{Cecil1988} and is roughly
consistent with the 690~\kms\ shock velocity inferred from the
\Chandra\ X-ray temperature \citep{Terashima2001}.

Cloud 1 lies at the northern edge of the XNC near the location where
the radio jet impinges on the radio lobe.  \cite{Cecil1988}
decomposed the XNC into several kinematic components using
ground-based ($\sim$1\arcsec\ resolution) Fabry-Perot observations at
65~\kms\ resolution, which is similar to our 62~\kms\ resolution at
\OIIIwb\ in G430M.  We note that his blueshifted XNC component ($v =
-180\ \kms,\ \Delta v = 140\ \kms$) is in close agreement with our
measurements for cloud 1 ($v = -145\ \kms,\ \Delta v = 104\ \kms$).

\subsection{Spectral Classification}

Based upon optical studies of the emission-line ratios, M51 has been
previously classified as a LINER \citep[e.g.][]{Stauffer1982,
Filippenko1985, Carrillo1999}, a Seyfert/LINER transition object
\citep{Heckman1980b}, and a Seyfert~2 \citep[e.g.][]{Ho1997a}.  Our
NLR cloud spectra exhibit strong emission lines of \OIIIww, \NIIww,
\OIIw, and \SIIww\ that are typical in Seyfert~2 spectra.

\begin{figure}[ht]
\epsscale{1.2}
\plotone{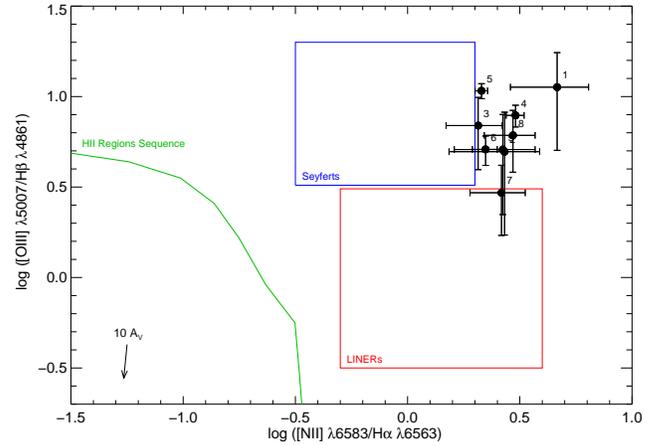}
\caption{\OIIIwb/\Hb\ vs. \NIIwb/\Ha\ diagnostic plot for the
emission-line clouds.  The loci of points typically occupied
by Seyferts and LINERs are denoted by the colored boxes.
The green line represents the theoretical \HII\ region track
of \cite{McCall1985}.  The arrow indicates the direction of the
reddening correction.}
\label{fig:vofig1}
\end{figure}

\begin{figure}[ht]
\epsscale{1.2}
\plotone{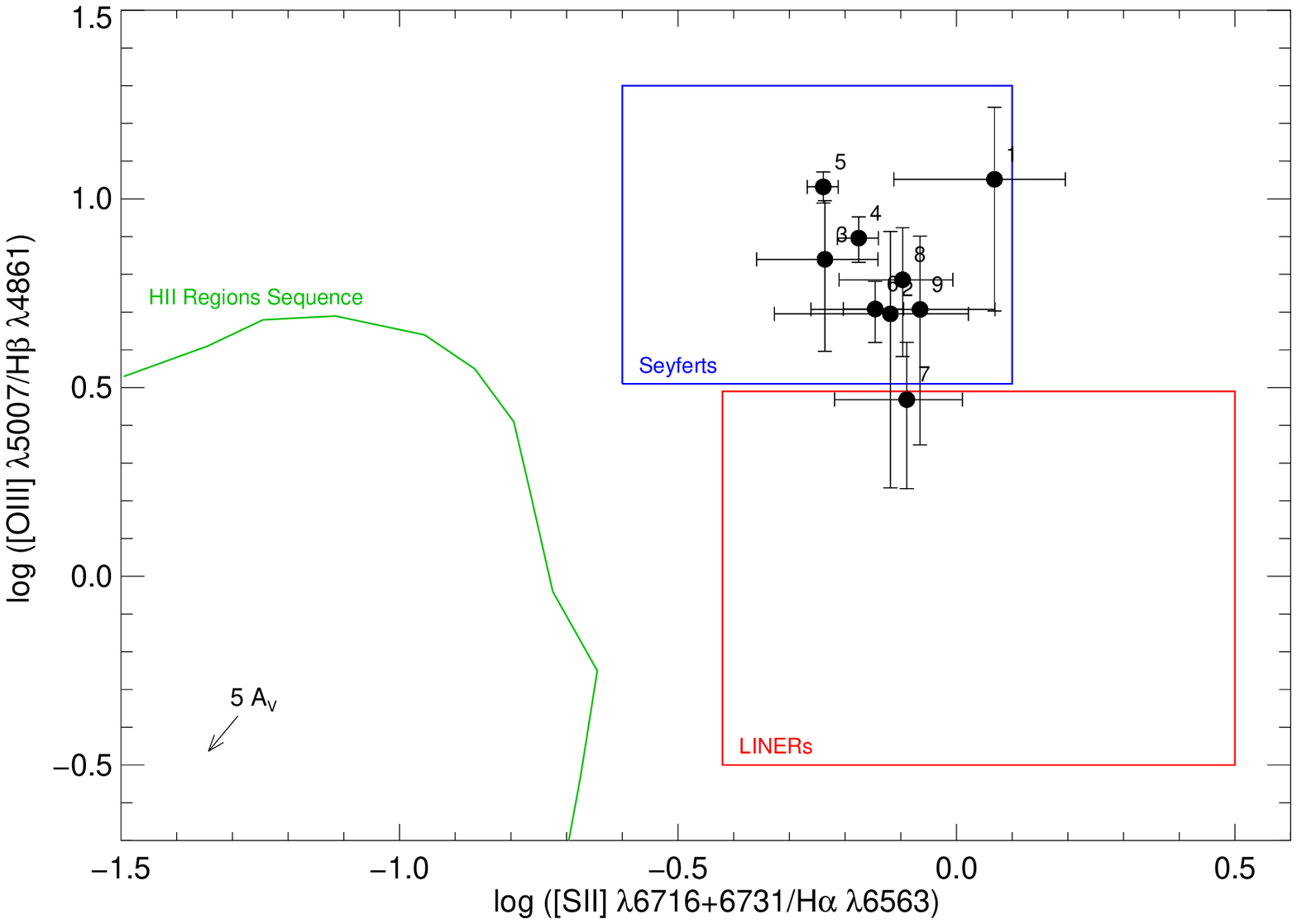}
\caption{As in Fig.~\ref{fig:vofig1}, but for \OIIIwb/\Hb\
vs. \SIIww/\Ha.}
\label{fig:vofig2}
\end{figure}

\begin{figure}[ht]
\epsscale{1.2}
\plotone{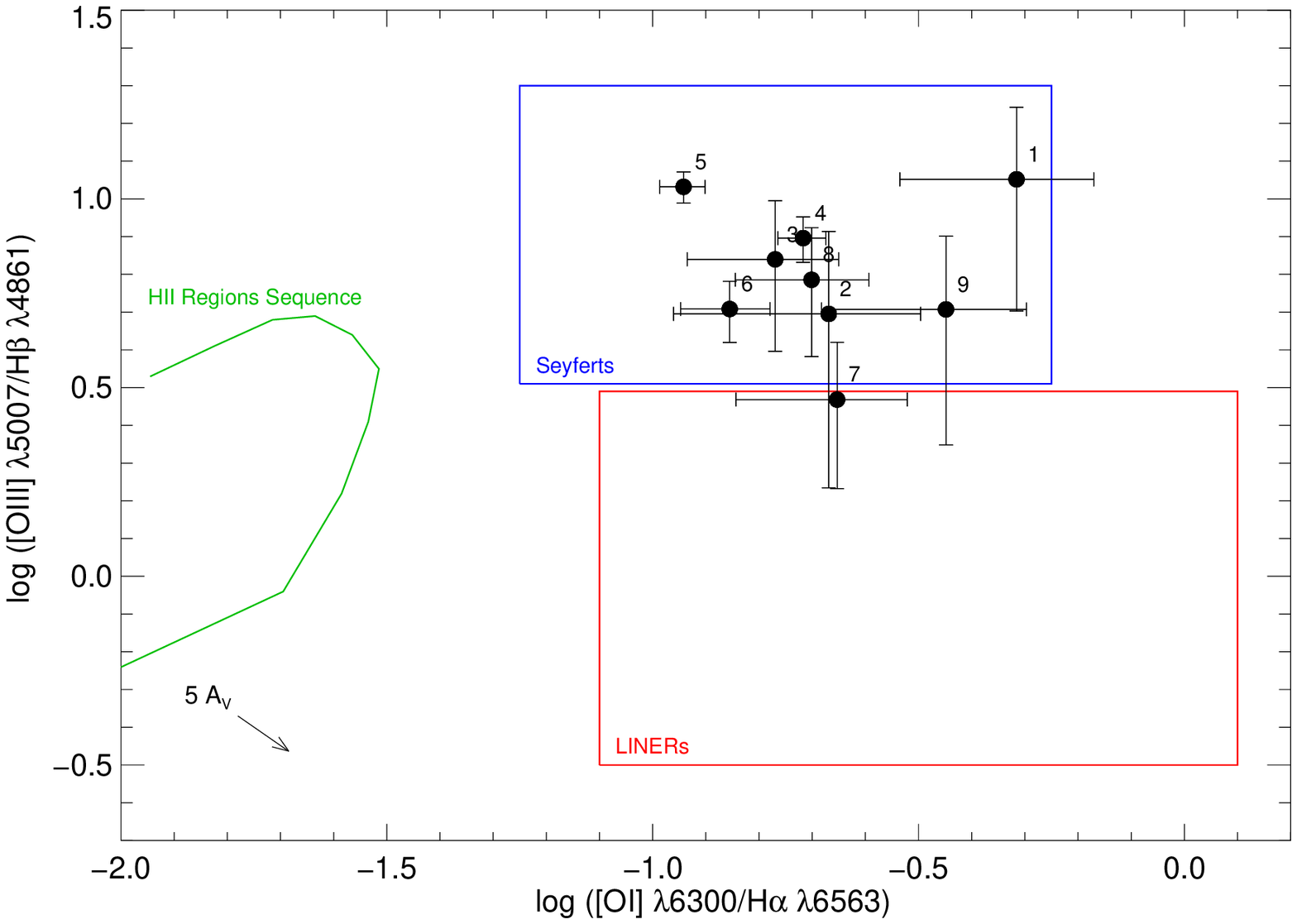}
\caption{As in Fig.~\ref{fig:vofig1}, but for \OIIIwb/\Hb\
vs. \OIwa/\Ha.}
\label{fig:vofig3}
\end{figure}

To classify emission-line objects, \cite*{Baldwin1981} used a series
of two-dimensional emission-line ratio diagnostic plots to
discriminate among various ionization mechanisms.  We adopted this
methodology and followed the convention of \citet[hereafter
VO]{Veilleux1987} who chose ratios of bright emission lines that are
close together in wavelength to minimize the impact of dereddening
errors.  Figures~\ref{fig:vofig1}, \ref{fig:vofig2}, and
\ref{fig:vofig3} show the three VO diagrams of \OIIIwb/\Hb\ versus
\NIIwb/\Ha, \SIIww/\Ha, and \OIwa/\Ha, respectively.

Because the partially-ionized transition zone in nebula photoionized
by young, hot stars is limited in size \citep{Stromgren1939},
\HII\ regions and starburst galaxies typically exhibit weaker
ionization lines than those observed in LINERs or Seyferts.  While
both LINERs and Seyferts are ionized by more energetic sources,
LINERs are distinguished from Seyferts by their lower excitation,
which is indicated in part by a lower value of \OIIIwb/\Hb\
\citep{Ho1993}.

In Figures~\ref{fig:vofig1}, \ref{fig:vofig2}, and \ref{fig:vofig3}
we denote the loci of points typically occupied by \HII\ regions,
LINERs, and Seyferts.  To represent the regions photoionized by OB
stars, which include \HII\ regions and starburst galaxies, we plotted
the theoretical \HII\ model sequence of \cite{McCall1985} for $T_{*}
= 38,500 - 47,000$~K.  The M51 emission-line ratios occupy regions
consistent with Seyfert galaxies.  In general we find higher
excitation, as measured by the \OIII/\Hb\ line ratios, for the NLR
clouds than is typically observed in LINER galaxies.  Cloud 7, our
weakest NLR cloud, has a somewhat lower value of \OIII/\Hb, but it
lies near the Seyfert/LINER boundary.

Figure~\ref{fig:vofig1} indicates that M51 possesses larger
\NIIwb/\Ha\ ratios ($\simeq 2.7$) than the region typically occupied
by Seyfert nuclei.  We believe this to be an abundance effect,
whereby nitrogen is selectively enhanced in the near-nuclear region
of M51 by factors of $\sim3-4.5$ (see \S~\ref{sec:modelresults}).

\subsection{Ionization State of the Gas}

\begin{figure*}[ht]
\epsscale{1.1}
\plotone{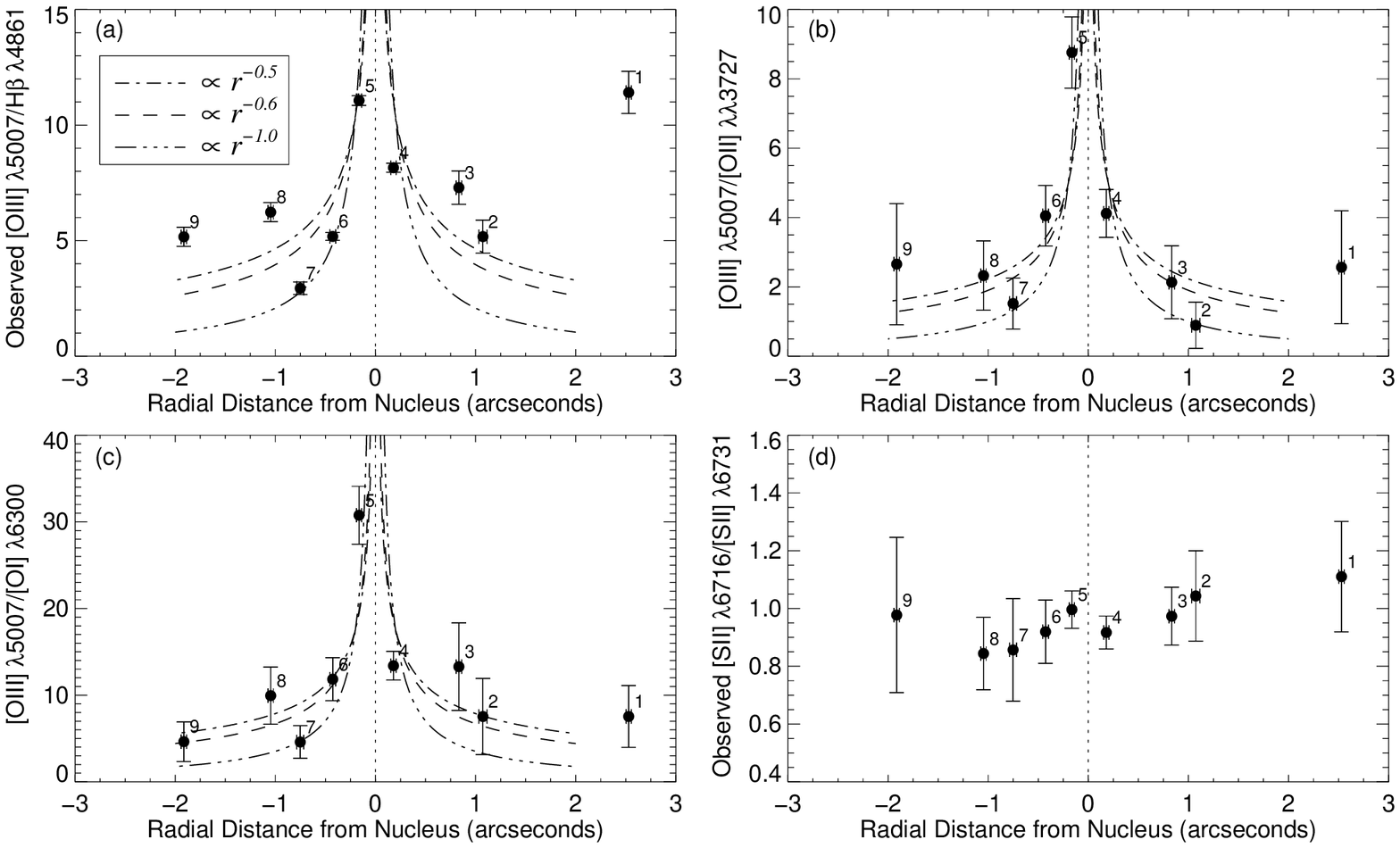}
\caption{(a) \OIIIwb/\Hb\ (b) \OIIIwb/\OIIw\ (c) \OIIIwb/\OIwa\
and (d) \SIIwa/\SIIwb\ for the nine NLR emission-line clouds
plotted as a function of radial distance from the nucleus.}
\label{fig:fluxratios}
\end{figure*}

The ionization state of the NLR gas can be examined through the use
of the \OIIIwb/\Hb\ ratio and the abundance insensitive
\OIIIwb/\OIIw\ and \OIIIwb/\OIwa\ line ratios.  The \OIIIwb/\OIIw\
line ratio is roughly proportional to the ionization parameter for
radiation-bounded clouds and is relatively independent of the shape
of the ionizing continuum \citep{Penston1990}.

Figures~\ref{fig:fluxratios}a$-$c show \OIIIwb/\Hb, \OIIIwb/\OIIw,
and \OIIIwb/\OIwa, respectively, as a function of projected radial
distance from the nucleus.  In general, these line ratios are larger
near the nucleus and decrease with increasing radius out to
$\sim$1\arcsec.  As a comparison, we have also overplotted three
$r^{-n}$ ($n = 0.5, 0.6, 1.0$) curves, which qualitatively agree with
the data, on Figures~\ref{fig:fluxratios}a$-$c.  The declining line
ratios indicate that the excitation of the NLR gas decreases with
increasing radius within the near-nuclear region.  As an exception,
cloud 1 possesses the largest value of \OIIIwb/\Hb\ (even larger than
the innermost clouds 4 and 5), which may indicate the presence of an
additional source of ionization (e.g. shocks) in this cloud.

\subsection{Electron Density \label{sec:eden}}

In the low-density regime, the \SIIww\ doublet is a well-known
electron density diagnostic that is relatively insensitive to
temperature \citep{Osterbrock1989}.  This diagnostic can be applied
in regions where the electron density is less than the critical
density of \SII\ ($3.3 \times 10^{3}$~\ivol).  Above this density the
lines become collisionally de-excited.  Unfortunately, the
\SIIww\ lines are severely blended in our data.  The wavelength
separation of these lines (14.4~\AA) is less than the spectral
resolution of the G750L grating mode (18.3~\AA).  We deblended the
\SIIww\ doublet using the \OIIIwb\ line as a template, however, the
resulting flux errors are large (Fig.~\ref{fig:fluxratios}d).  We
note that for a \SII\ $\lambda6716/\lambda6731$ ratio near unity, a
10\% error in each of the emission lines results in nearly a factor
of two uncertainty in the electron density.

To calculate the electron density of each NLR cloud
(Table~\ref{tbl:conditions}), we used the ``temden'' routine of the
IRAF NEBULAR package, which uses a five-level atom approximation
\citep{Shaw1995}.  For clouds 2$-$9, we assumed an electron
temperature $T_{e} = 10^4$~K, which is typical for photoionized gas
in the NLR of Seyfert galaxies.  For cloud 1, we used an electron
temperature of $2.4 \times 10^4$~K, which was calculated from the
observed \OIII\ emission lines (see \S~\ref{sec:etemp}).  The
resulting electron densities lie in the range between 490~\ivol\ and
1150~\ivol, however, no trend in the gas density as a function of
radius can be inferred given the large flux errors.  The average
\SII\ electron density within the NLR gas is 770~\ivol.  Cloud 1,
located at the largest radial distance (2\farcs5), possesses the
lowest electron density of $n_{e}=490$~\ivol.

\begin{deluxetable*}{cccccc}
\tabletypesize{\normalsize}
\tablecolumns{6}
\tablewidth{0pt}
\tablecaption{Physical Conditions in the NLR Clouds Along P.A. 166\dg \label{tbl:conditions}}
\tablehead{\colhead{Cloud}
   &\colhead{Radius\tablenotemark{a}} 
   &\colhead{$n_{e}$}
   &\colhead{$T_{e}$}
   &\colhead{$L(\Hb)$}
   &\colhead{Ionized Mass} \\
 & \multicolumn{1}{c}{(\arcsec, pc)} & \multicolumn{1}{c}{(\ivol)} & \multicolumn{1}{c}{(K)} & \multicolumn{1}{c}{($10^{36}$~\ergs)} & \multicolumn{1}{c}{($M_{\odot}$)} }
\startdata
1  & \phs2.53 (103)    & \phn490 & \phantom{$<$} 24,000 & \phn2.7 & \phnn50 \\
2  & \phs1.07 \phn(44) & \phn510 & $<$ 21,800           & \phn3.0 & \phnn60 \\
3  & \phs0.83 \phn(34) & \phn680 & $<$ 17,500           & \phn6.5 & \phnn90 \\
4  & \phs0.18 \phnn(7) & \phn850 & $<$ 10,200           &    31.1 & \phn350 \\
5  &  $-$0.16 \phnn(7) & \phn620 & $<$ \phn9,900        &    30.2 & \phn470 \\
6  &  $-$0.43 \phn(18) & \phn840 & $<$ 11,000           & \phn9.8 & \phn110 \\
7  &  $-$0.75 \phn(31) &    1090 & $<$ 19,200           & \phn2.0 & \phnn20 \\
8  &  $-$1.05 \phn(43) &    1150 & $<$ 11,100           & \phn3.7 & \phnn30 \\
9  &  $-$1.91 \phn(78) & \phn670 & $<$ 15,800           & \phn1.7 & \phnn30 \\
Total & \nodata        & \nodata & \nodata              &    90.7 &    1210 \\[-1em]
\enddata 
\tablenotetext{a}{Positive values denote clouds south of the nucleus.  Negative values denote clouds north of the nucleus.}
\end{deluxetable*}

We compare our results with those of \cite{Rose1982} who adopted an
average value of \SII\ $\lambda6716/\lambda6731 = 1.0$ for the entire
nuclear region of M51 (\slit{61\farcs4}{2\arcsec} slit).  These
authors assumed $T_{e} = 8 \times 10^{3}$~K and calculated an
electron density of 630~\ivol.  If we adopt this electron
temperature, we find an average electron density of 690~\ivol, which
is in very good agreement with their value.  Further, our results are
similar to those of \cite{Rose1983} who found in a 5\arcsec\ circular
aperture an average \SII\ $\lambda6716/\lambda6731 = 0.93$,
corresponding to an electron density of 880~\ivol\ for $T_{e} =
7620$~K.  Our calculated electron densities are also consistent with
those found by \cite{Ford1985} who obtained electron densities in the
range 640$-$720~\ivol\ for the nucleus of M51 and 370$-$490~\ivol\
for the XNC.

\subsection{Electron Temperature \label{sec:etemp}}

The electron temperature can be estimated from the observed \OIII\
($\lambda4959\,+\,\lambda5007$)/$\lambda4363$ ratio.  The difficulty
in using this diagnostic is that the \OIIIwc\ emission line generally
is very weak in Seyfert spectra.  For our NLR cloud spectra, we
observed \OIIIwc\ only in cloud 1.  Using the NEBULAR package, we
calculated an electron temperature of 24,000~K for this cloud
(assuming an electron density of 490~\ivol).  Electron temperatures
of this scale typically are not observed in the O$^{+2}$ zone of
photoionized gas \citep{Kraemer1998}, providing further evidence that
shocks contribute to the ionization structure of this cloud.

Because the \OIIIwc\ line was too weak to be detected in the other
NLR clouds, we derived upper limits for the electron temperature of
the other clouds by numerically integrating the flux over the
strongest noise feature in the spectrum near 4363~\AA\
(Table~\ref{tbl:conditions}).  The resulting electron temperature is
$\la$~11,000~K for the three innermost clouds (4$-6$) and cloud 8.
These temperatures effectively rule out collisional ionization as the
dominant source of excitation in these clouds as electron
temperatures $T_e\ < 20,000$~K are too low for this mechanism to be
important \citep{Wilson1993b}.

Clouds 3 and 9 have electron temperatures $T_e\ \la$ 17,500~K, which
also suggests that the ionization of these clouds is primarily due to
photoionization.  The upper limits for the temperature of clouds 7
and 2 are 19,000~K and 21,800~K, respectively.  It is most likely
that these represent weak constraints and the large upper limits are
due to the lower signal-to-noise of these clouds, which are fainter
than the inner NLR clouds and cloud 1.  While it remains that the
larger upper-limit to the electron temperatures could possibly
indicate a greater role for shocks, the MAPPINGS~II shock models (see
\S~\ref{sec:shocks}) provide a poorer fit to these clouds than our
photoionization models.

\subsection{Ionized Mass}

The mass of ionized gas in each NLR cloud can be estimated from the
observed electron density and luminosity of the \Hb\ recombination
line.  For a completely ionized gas with solar helium abundances, the
total ionized mass is given as \citep{Osterbrock1989}
\begin{equation}
M\ =\ \frac{1.4\ m_{H}\, 
L(\mbox{H}\beta)}{\alpha_{H\beta}^{eff}\ n_e\, h\nu_{H\beta}} \ ,
\end{equation}

\noindent
where $m_{H}$ is the mass of a hydrogen atom, $L(\Hb)$ is the
observed luminosity (\ergs) of the \Hb\ recombination line,
$\alpha_{H\beta}^{eff}$ is the Case~B effective \Hb\ recombination
coefficient \citep[$3 \times 10^{-14}$~cm$^3$~s$^{-1}$ at
$10^{4}$~K,][]{Osterbrock1989}, $n_e$ is the electron density, and
$h\nu_{H\beta}$ is the energy of an \Hb\ photon.  We derived the
\Hb\ luminosities by correcting the observed \Hb\ emission-line
fluxes for the distance to M51 of 8.4~Mpc \citep{Feldmeier1997}.  The
electron densities were taken as those derived in \S~\ref{sec:eden}
for the \SII\ emitting region.  The resulting ionized masses of each
NLR cloud are shown in Table~\ref{tbl:conditions}.  The total ionized
mass of the NLR clouds in our \slit{52\arcsec}{0\farcs2}\ slit is
$1.2 \times 10^{3}\ M_{\odot}$, with most of this mass located in the
three innermost and brightest \Hb\ clouds.  For comparison,
\cite{Ford1985} derived a total ionized mass of $4.4 \times
10^{3}\ M_{\odot}$ within a larger \slit{2\arcsec}{2\arcsec} region
centered on the nucleus.  While our slit position was chosen to
intersect the bright \OIII\ clouds straddling the nucleus and the
bright NLR cloud located near the jet terminus, additional
structures, such as two bright clouds north of nucleus and faint
extended circumnuclear emission, are included in Ford's measurement.

\subsection{Ionizing Flux}

The observed \Hb\ luminosity can also be used to calculate the total
number of recombinations per second as \citep{Peterson1997}
\begin{equation}
Q(H)\ =\ \frac{L(\mbox{H}\beta)}{h \nu_{H\beta}} \frac{\alpha_{B}}{\alpha_{H\beta}^{eff}}\ =\ 2.11 \times 10^{12}\ L(\mbox{H}\beta)\ \ \mbox{photons~s}^{-1} \ ,
\end{equation}

\noindent
where $\alpha_{B}$ is the Case~B hydrogen recombination coefficient
\citep[$2.6 \times 10^{-13}$~cm$^{3}$~s$^{-1}$ at
$10^{4}$~K,][]{Osterbrock1989}.

For the nine NLR clouds located within our
\slit{52\arcsec}{0\farcs2}\ slit along position angle 166\dg, the
total \Hb\ luminosity is $9.1 \times 10^{37}$~\ergs\ (see
Table~\ref{tbl:conditions}).  Using the above equation, we derive a
hydrogen recombination rate of $Q(H) = 1.9 \times
10^{50}$~photons~s$^{-1}$.  This number provides an estimate of the
total number of hydrogen ionizing photons produced per second by the
active nucleus by assuming ionization equilibrium, whereby the total
number of hydrogen ionizing photons produced by the central source
equals the total number of recombinations.

Because the observed \Hb\ luminosity was summed over our single slit
position, we have not accounted for all of the NLR \Hb\ emission.  As
such, $Q(H)$ represents a lower-limit to the rate of ionizing photons
produced by the central source.  Within a \aslit{2}{2} aperture,
which as discussed previously contains emission structures not
included in our narrow slit, \cite{Ford1985} reported a nuclear
\Hb\ luminosity of $2.8 \times 10^{38}$~\ergs.  This results in a
total rate of hydrogen ionizing photons of $5.9 \times
10^{50}$~photons~s$^{-1}$.  These calculations show the relative
weakness of the active nucleus of M51.  For comparison,
\cite{Schulz1993} derived a total rate of hydrogen ionizing photons
of $3 \times 10^{54}$~photons~s$^{-1}$ for the central source of the
luminous Seyfert~1 galaxy NGC~4151.

\section{Radio Observations}
\label{sec:radioobs}

The existence of a radio jet and radio lobe on the southern side of
the nucleus of M51 was first detected using VLA observations at
4.8~GHz \citep{Crane1992}.  These authors found a sinuous radio jet,
extending $\sim$3\arcsec, that connected the weak nuclear radio
source with a diffuse radio lobe, which has been identified with the
optical XNC feature \citep{Ford1985}.  The radio lobe was dominated
by a bright arcuate, ridge-like structure extending approximately
5\arcsec.

Our VLA-A 8.4~GHz observations (see Fig.~\ref{fig:radiomaps}) confirm
the presence of the previously observed radio structures in the
near-nuclear region of M51.  The radio maps show a weak, unresolved
radio core with a flux density of 330~\muJybm.  The southern radio
jet spans 2\farcs3 (94~pc) and decreases in luminosity along its
spatial extent, reaching a minimum of $\sim$2$-$4~\muJybm.  The
position angle at the base of the jet is 158\dg.  The radio jet
terminates near a diffuse, extended lobe structure, however, it is
not connected to the radio lobe at the lowest contour level.  It
appears that the radio jet connects to the southern radio lobe near
the location of cloud 1, however, because of the low signal-to-noise
of our radio observations we are unable to determine the precise
location of the impingement.  Although our resolution at 3.6~cm is
better than the earlier 6~cm data \citep{Crane1992}, a comparison of
the peak flux in the jet at these two wavelengths suggests that the
spectral index is approximately 3.1.  The total east-west extent of
the southern radio lobe is 5\farcs3 (220~pc).

The nuclear source of M51 is also surrounded by a 2\arcsec\ extended
radio structure that is elongated along a position angle of 169\dg.
To the north-northwest of the nucleus, this structure turns into a
ridge that extends approximately 1\farcs5 and could signify a bent
northern counter-jet.  Our 3.6~cm data do not exhibit the northern
loop structure observed at 20~cm by \cite{Ford1985} and at 6~cm by
\cite{Crane1992}.  Because this feature is only weakly detected at
6~cm, it is likely not present in our maps due to the lower
signal-to-noise of our radio data and is probably not a result of the
frequency difference.

In addition to the nuclear source structures, we find an elongated
triple radio source at 27\farcs8 (1130~pc) north-northwest of the
nucleus at a position angle of 335\dg.  At the distance of M51, the
spatial length of this structure is 86~parsec.  The position angle of
this northern source is within 14\dg\ of the position angle of the
nuclear elongation, and it is possible that this triple structure is
a remnant of past jet activity.  However, it could also result from a
weak radio galaxy in the background.

The centerline of the radio jet and its brightness distribution
reveal a substantial non-linearity that is suggestive of bends and
knots in the jet.  We identify eleven radio knots along the entire
north-south structure.  The location and the flux density of these
radio knots are presented in Table~\ref{tbl:radopt}.  The earlier
4.8~GHz radio maps only displayed two distinct hotspots.

\begin{deluxetable*}{lcccccl}
\tabletypesize{\footnotesize}
\tablecolumns{7}
\tablewidth{0pt}
\tablecaption{Radio-Optical Correlations\label{tbl:radopt}}
\tablehead{\colhead{Radio Cloud}
   & \colhead{Radius\tablenotemark{a}}
   & \colhead{Radio Flux}
   & \colhead{Optical Cloud} 
   & \colhead{Radius\tablenotemark{a}}
   & \colhead{\Hb\ Flux} 
   & \colhead{Comments} \\
   & \multicolumn{1}{c}{(\arcsec, pc)} 
   & \multicolumn{1}{c}{($\mu$Jy/beam)} 
   & 
   & \multicolumn{1}{c}{(\arcsec, pc)} 
   & \multicolumn{1}{c}{($10^{-16}$ \ergscm)}
   &
}
\startdata
\sidehead{North}
R10     &  $-$1.8 \phn(73) & 2.6 & \nodata &     \nodata       &  \nodata & Adjacent to Cloud 9 \\
R9      &  $-$1.2 \phn(49) & 2.2 & Cloud 8 &  $-$1.05 \phn(43) & \phn4.43 & Adjacent to Cloud 8 \\
R8      &  $-$0.8 \phn(33) & 1.6 & Cloud 7 &  $-$0.75 \phn(31) & \phn2.32 & Adjacent to Cloud 8 \\
R7      &  $-$0.6 \phn(24) & 2.8 & Cloud 6 &  $-$0.43 \phn(18) &    11.62 & Adjacent to Cloud 6 \\
R6      &  $-$0.3 \phn(12) & 3.6 & Cloud 5 &  $-$0.16 \phnn(7) &    35.73 & Adjacent to Clouds 5 \& 6 \\[2ex]
Nucleus &      0           & 155 & \nodata &      0            &  \nodata & \\
\sidehead{South}
R5      & \phs0.6 \phn(24) & 3.6 & Cloud 3 & \phs0.83 \phn(34) & \phn7.66 & Cloud 3 at edge of R5 \\
R4      & \phs1.0 \phn(41) & 1.6 & Cloud 2 & \phs1.07 \phn(44) & \phn3.54 & See morphology note\tablenotemark{b} \\
R3      & \phs1.4 \phn(57) & 2.6 & \nodata &     \nodata       &  \nodata & See morphology note\tablenotemark{b} \\
R2      & \phs1.7 \phn(69) & 2.6 & \nodata &     \nodata       &  \nodata & See morphology note\tablenotemark{b} \\
R1      & \phs2.3 \phn(94) & 3.1 & Cloud 1 & \phs2.53 (103)    & \phn3.16 & R0 \& R1 straddle Cloud 1 \\
R0      & \phs2.7 (110)    & 2.6 & Cloud 1 & \phs2.53 (103)    & \phn3.16 & R0 \& R1 straddle Cloud 1
\enddata
\tablenotetext{a}{Positive values denote clouds south of the nucleus.  Negative
values denote clouds north of the nucleus.}
\tablenotetext{b}{Morphologically, the region between R4 and R1
appears as though it may have been evacuated by the radio jet.}
\end{deluxetable*}

Our VLA 8.4~GHz data reveal a slightly different jet orientation as
compared to the earlier low-resolution 4.8~GHz data
\citep{Crane1992}.  Specifically, the position angle of the southern
radio jet in our 8.4~GHz maps (observed in 1999) lies $\sim$10\dg\ to
the east of the centerline of the earlier 4.8~GHz jet structure
(observed in 1985/1986).  Additionally, the brightness distribution
of the jet and southern radio lobe structure appears different than
the earlier observations.  In particular, the southern lobe structure
is rather patchy and does not exhibit the clear arcuate,
emission-ridge structure that was observed in the earlier 4.8~GHz
data.

A potentially intriguing explanation for the discrepancy in the radio
structures in the south would be oscillatory structure in the jet.
The total radio power in the nuclear region of M51 ($2 \times
10^{19}$~W~Hz$^{-1}$) is at the very bottom of the distribution of
active galaxies \citep{Condon1992}, and the relative weakness of the
radio jet makes it susceptible to instabilities that can cause the
jet to precess.  Oscillatory structure may be the result of
precession at the nucleus itself or from Kelvin-Helmholtz (KH)
instabilities resulting from the velocity differences between the gas
inside and outside the jet.  The KH instability results in helical
structure for the jet that shows an exponentially growing oscillation
with distance along the jet.  However, the deduced shift in position
angle of a precessing jet is relatively large and would result in a
precession period that is too short.  The mis-orientation of the low-
and high-resolution jets could be due to a limb-brightening effect,
but this would require rather unrealistic frequency-dependent opacity
variations across the width of the jet.  Alternatively, the apparent
shift could simply result from the signal-to-noise limitations of our
data and the uniformly weighted map of \cite{Crane1992}.

\begin{figure*}[ht]
\epsscale{0.98}
\plotone{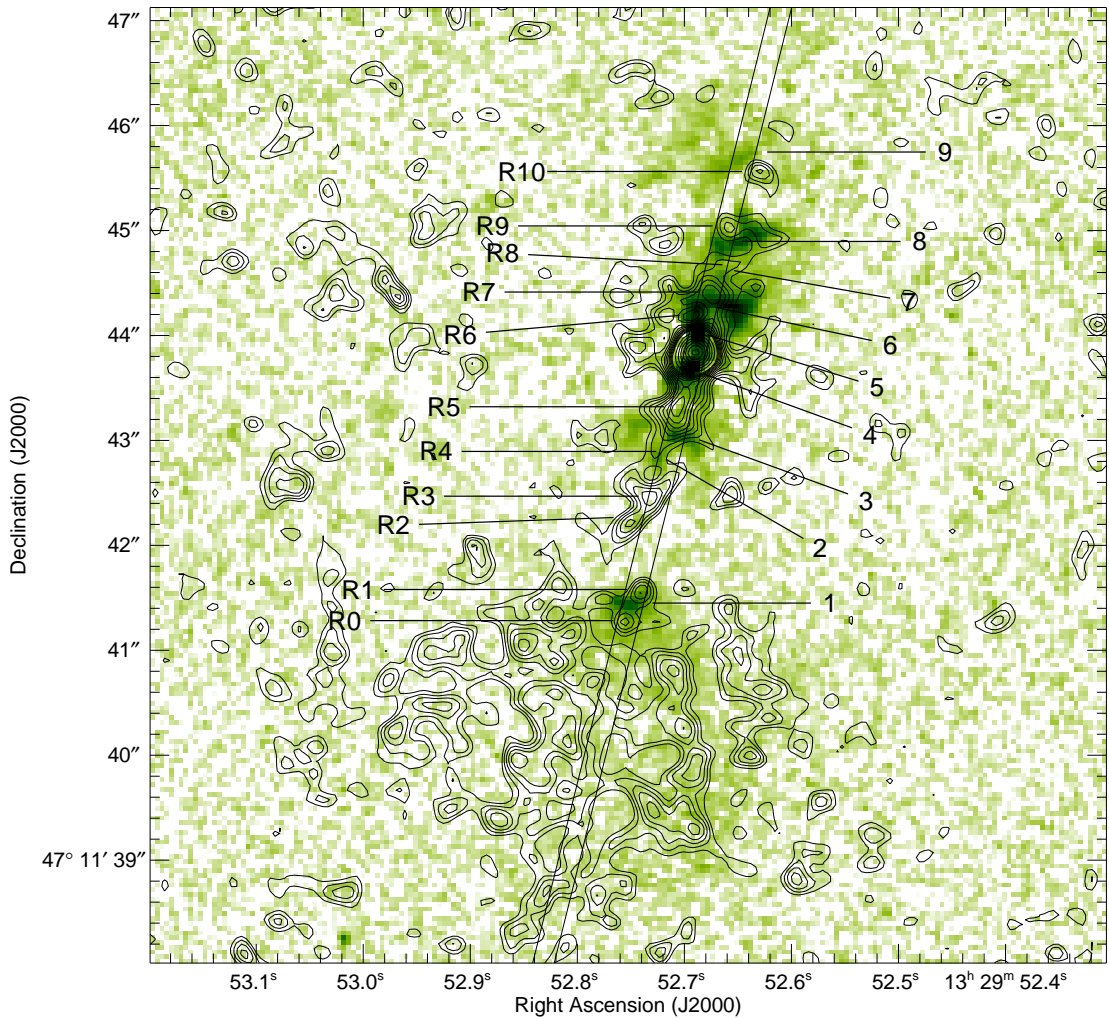}
\caption{Continuum subtracted \HST/WFPC2 F502N narrowband \OIII\
image superimposed with our VLA 8.4~GHz radio contours.  The contour
levels are [2,3,4,5,6,7,8,9,10,11,12,16,20,24,28,32,36,40] $\times$
5.0~$\mu$Jy/beam.  The location of the optical NLR clouds and radio
knots have been indicated as well as the position of the STIS
\slit{52\arcsec}{0\farcs2} slit.}
\label{fig:f502nradio}
\end{figure*}

\begin{figure*}[ht]
\epsscale{0.98}
\plotone{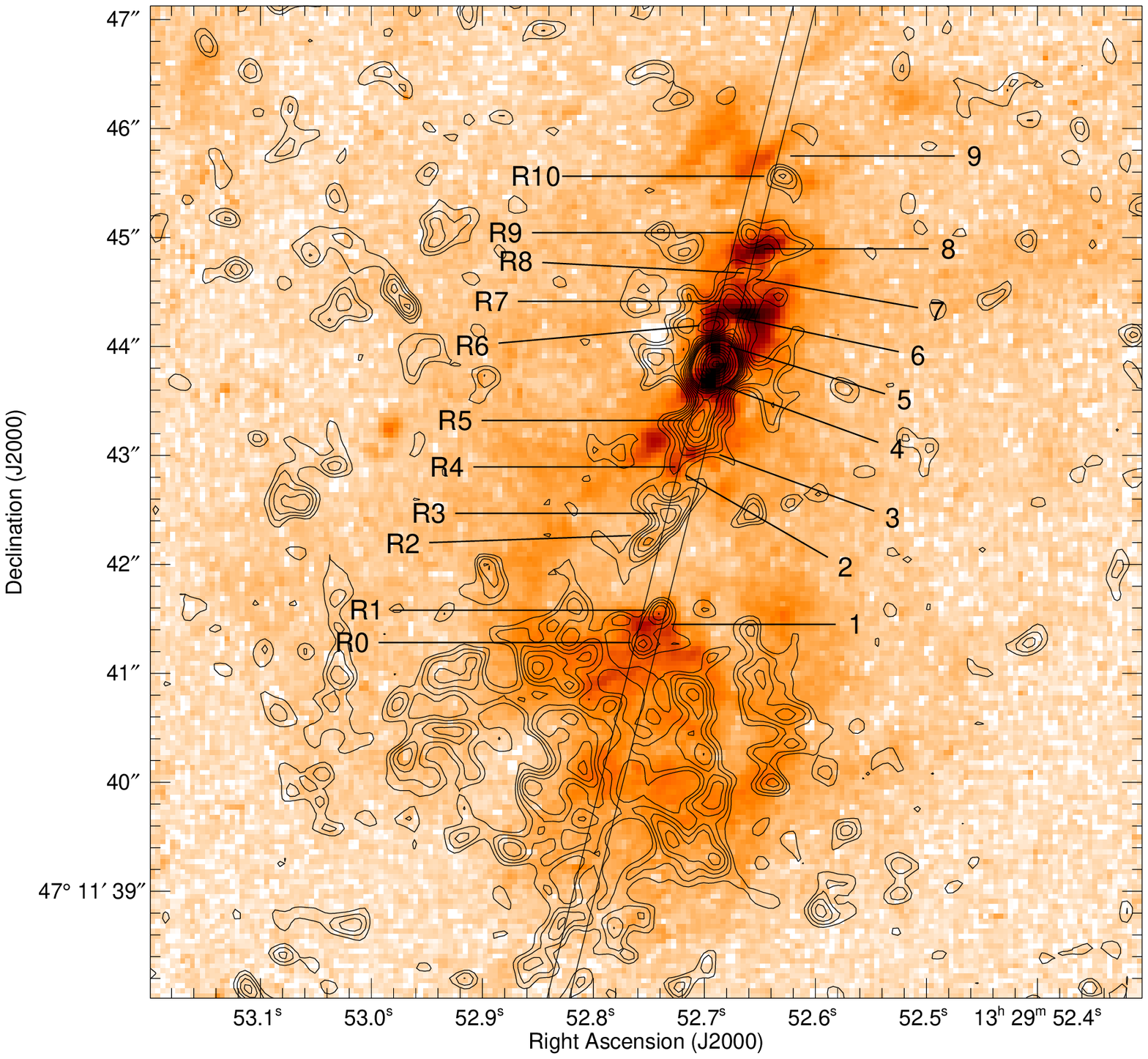}
\caption{Continuum subtracted \HST/WFPC2 F656N narrowband
$\NII\,+\,\Ha$ image superimposed with our VLA 8.4~GHz radio
contours.  The contour levels are
[2,3,4,5,6,7,8,9,10,11,12,16,20,24,28,32,36,40] $\times$
5.0~$\mu$Jy/beam.  The location of the optical NLR clouds and radio
knots have been indicated as well as the position of the STIS
\slit{52\arcsec}{0\farcs2} slit.}
\label{fig:f656nradio}
\end{figure*}

In Figures~\ref{fig:f502nradio} and \ref{fig:f656nradio}, we present
the \HST/WFPC2 \OIII\ and $\NII\,+\,\Ha$ narrowband images superposed
with the VLA 8.4~GHz radio continuum contours.  In order to align the
radio core with the assumed location of the obscured nucleus, the
radio image (\slit{0\farcs29}{0\farcs26} resolution) was shifted
relative to the optical images (0\farcs09 resolution) by $\Delta
\alpha=0\farcs15$ and $\Delta \delta=-1\farcs06$.  The location of
the nine optical emission-line clouds, the eleven radio knots, and
the slit for the STIS spectra also have been indicated.

The radio structures are elongated with a position angle close to
that of the optical NLR emission observed in the \HST\ images.  The
STIS \slit{52\arcsec}{0\farcs2} slit falls on most of the inner part
of the elongated nuclear radio structure.  Six of the nine optical
NLR clouds located within our slit lie near the projected position of
radio features.  At the nucleus, the extended $\NII\,+\,\Ha$ and
\OIII\ emission falls within the outer radio contours.  The overall
morphological agreement of the optical and radio emission at 8.4~GHz
is very suggestive.  We find that the optical emission-line clouds,
in general, lie adjacent to the peaks of the radio emission.  This
spatial anti-coincidence suggests that the radio jet and its
structural components do affect the optical emission regions and are
associated as an energy source with the optical emission-line
clouds.  However, there are a couple of radio knots within the radio
jet/optical bicone region that do not have nearby optical knots and
thus, not all of the radio knots are anti-correlated with an
emission-line cloud.  While the radio resolution is $\sim3$ times
worse than the optical image, it appears that the radio and optical
knot anti-coincidence would still be present in a higher resolution
radio map.  South of the nucleus, it appears that the radio jet may
have evacuated a conical region (see Fig.~\ref{fig:f656nradio}).  In
Table~\ref{tbl:radopt} we provide a more detailed comparison of the
locations and properties of the radio knots with the \Hb\ emission
from the optical clouds.

The \OIII\ and $\NII\,+\,\Ha$ \HST\ images show that cloud 1, located
2\farcs5 south of the nucleus, falls within the radio lobe at the
northern edge and coincides with a clear radio hotspot at location
R0$-$R1.  This hotspot could be interpreted as the entry point of the
jet into the radio lobe.  However, earlier observations by
\cite{Cecil1988} associate higher velocity dispersion gas east of
this point with the entry point of the jet.  Higher signal-to-noise
radio observations are necessary to confirm the precise location at
which the radio jet impinges the radio lobe.

The WFPC2 F656N $\NII\,+\,\Ha$ image shows an extended shell-like
structure that surrounds the region where the radio jet is currently
passing through.  This shell is particularly prominent on the eastern
side of the source, while it is fragmented on the western side.  It
is intriguing to interpret this shell structure as the boundary of a
cavity region that has been evacuated by the radio jet.  The
excitation of this shell structure could be due to radiation emerging
from the energy dissipation within the jet.  Alternatively, the
excitation of the shell could result from lateral expansion shocks
generated by the twisting jet.

\section{Photoionization Models}
\label{sec:photomodels}

We have generated photoionization models to determine if the NLR
emission-line spectra are consistent with photoionization by a
non-thermal central continuum source.  The photoionization models
were computed using version 94 of the photoionization code Cloudy
\citep{Ferland1998}.  The models were parameterized by the shape of
the ionizing continuum, the dimensionless ionization parameter $U$,
and the total hydrogen density $n_{H}$.  The dimensionless ionization
parameter $U$ is defined as
\begin{equation} 
U=\frac{1}{4 \pi r^2 n_H c} \int_{\nu_{0}}^{\infty} \frac{L_{\nu}}{h \nu} 
\, d\nu \ ,
\end{equation}

\noindent
where $r$ is the distance from the inner (illuminated) face of the
cloud to the central continuum source, $n_H$ is the total hydrogen
density, $h\nu_{0}$ is the ionization energy of hydrogen (13.6~eV),
and $L_{\nu}$ is the frequency-dependent luminosity of the nuclear
ionizing continuum.  The models assume a constant density cloud with
a plane-parallel geometry and emission-line photon escape from the
illuminated face of the cloud.

\subsection{Model Input Parameters}

\begin{figure}[ht]
\epsscale{1.2}
\plotone{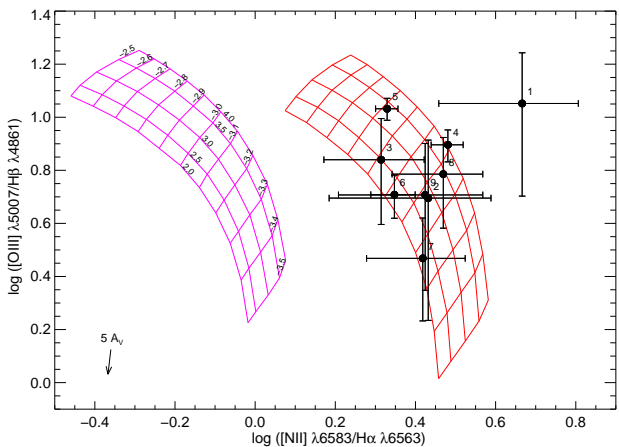}
\caption{\OIIIwb/\Hb\ vs. \NIIwb/\Ha\ for the NLR clouds overplotted
with our Cloudy photoionization model grids.  The solar abundance
model grid is shown in magenta and is labeled with $\log\,U=-2.5$ to
$-$3.5 and $\log\,n_{H}=2$ to 4~\ivol.  The red grid shows the
photoionization models with a $4\times$ enhancement of the nitrogen
abundance.}
\label{fig:mapvo1}
\end{figure}

Photoionization models with solar abundances cannot reproduce the
relatively strong observed \NIw\ and \NIIww\ nitrogen lines in M51.
We demonstrate this graphically in Fig.~\ref{fig:mapvo1}, where we
have plotted the photoionization model grids for both solar and four
times solar nitrogen (4~N$_{\odot}$).  The best-fitting models
require selective enhancement of nitrogen in the range of 3$-$4.5
times solar.  The numerical abundances relative to hydrogen for a
model with a factor of three overabundance of nitrogen are:
He~$=\xten{1.0}{-1}$, C~$=\xten{3.55}{-4}$, N~$=\xten{2.8}{-4}$,
O~$=\xten{7.4}{-4}$, Ne~$=\xten{1.17}{-4}$, S~$=\xten{1.62}{-5}$, and
Ar~$=\xten{3.98}{-6}$ \citep{Grevesse1989}.

We also generated photoionization models that included the presence
of dust grains within the line-emitting region.  However, when
compared to our observed optical emission lines, these models are
very similar to models without dust grains.  Ultraviolet spectra
and/or emission lines involving elements commonly found in dust
grains can be useful for constraining models with dust.  The IUE
spectrum (\aslit{10}{20} elliptical aperture) of the nuclear region
in M51 only shows the presence of a blended \semiforbw{Si}{3}{1892}
and \semiforbw{C}{3}{1909} emission line attributed to \HII\ regions
\citep{Ellis1982}.  We investigated whether the absence of
\CIVw\ from the {\em IUE} spectrum could be used to constrain dust
models.  For our highest excitation photoionization model, the
\CIVw\ reddening-corrected line flux ($\sim 1 \times
10^{-15}$~\ergscm) would be undetectable in the M51 {\em IUE}
spectrum.  Therefore without a more sensitive UV spectrum, we are
unable to constrain models with dust grains.  While the
photoionization models that follow do not include the presence of
dust grains within the NLR line-emitting region, we cannot rule out
such models.

Using published M51 continuum flux measurements, we constructed a
broadband spectral energy distribution (SED) to define the shape of
the ionizing continuum.  We assumed the SED was a power-law of the
form $F_{\nu}$ $\propto$ $\nu^{-\alpha}$, where the spectral index,
$\alpha$, was defined piecewise for different frequency intervals.
Numerous photoionization models were generated to investigate the
effect of changing the SED slope in all regions of the spectrum, in
particular the UV to X-ray wavelength regime.  As one would expect,
the models were most sensitive to the spectral index in the extreme
ultraviolet (EUV) region of the spectrum.  We started with
$\alpha_{EUV}=1.29$ ($h\nu\ \ga\ 13.6$~eV), which was derived from
the published data, and constructed several photoionization model
grids by varying $\alpha_{EUV}$ over the range 1$-$2.  We found the
best-fitting photoionization models occurred for
$\alpha_{EUV}=1.35$.  In addition, we tested a model with the
presence of an ultraviolet blue-bump in the ionizing continuum.
However, this model produced a poorer fit to our observed line
fluxes, especially for \NIw\ and \OIww.

The photoionization models were generated using Cloudy's optimize
command, where we allowed the ionization parameter, $U$, and total
hydrogen density, $n_{H}$, to vary for each cloud.  As previously
discussed, our early photoionization model grids without optimization
indicated that enhanced nitrogen abundances were required to fit the
observed emission-line spectra.  Therefore, the relative nitrogen
abundance also was allowed to vary in the modelling procedure.

\subsection{Model Results \label{sec:modelresults}}

\begin{figure*}[ht]
\epsscale{1.18}
\plotone{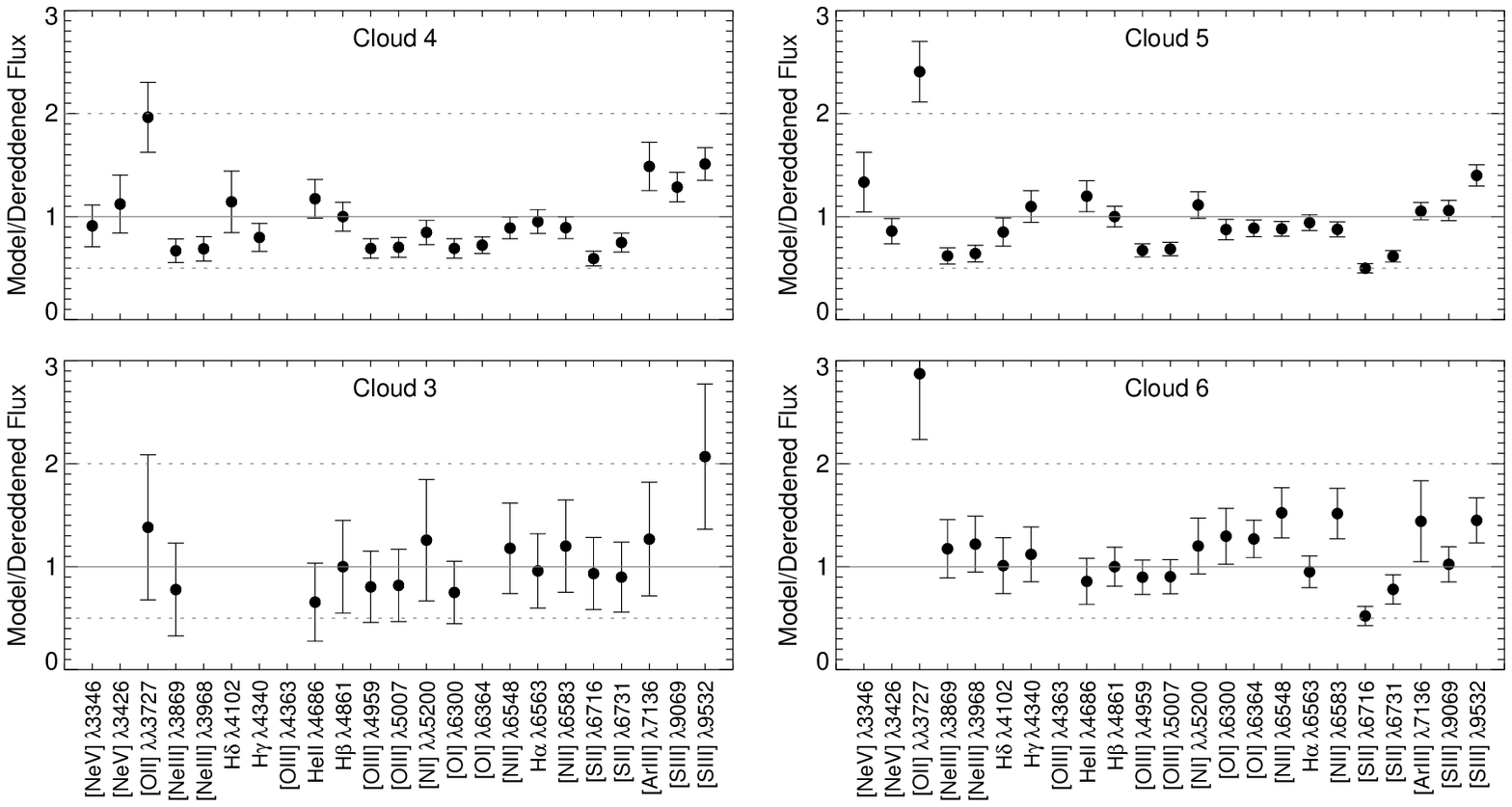}
\plotone{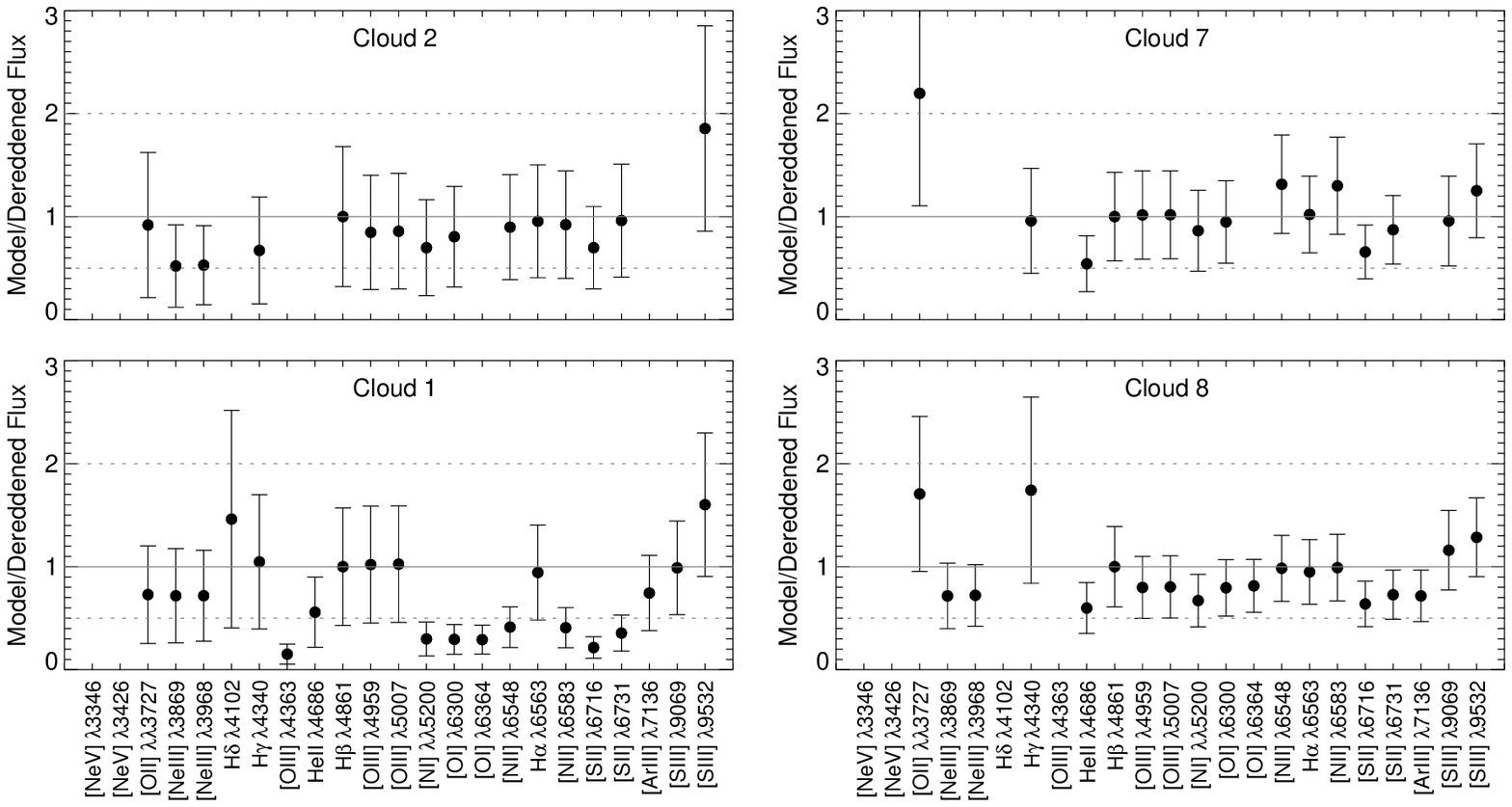}
\plotone{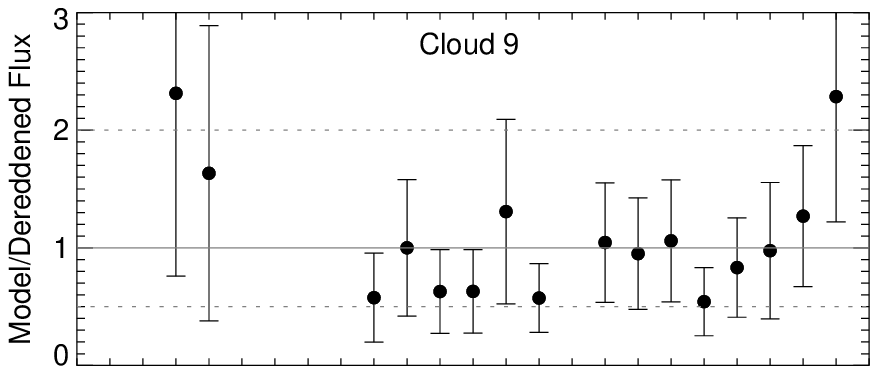}
\caption{\footnotesize 
Ratios of the photoionization model predictions to the
dereddened emission-line fluxes for each NLR cloud.  The left-hand
plots are for the NLR clouds south of nucleus, while those on the
right are for the NLR clouds north of the nucleus.  The radial
distance from the nucleus increases as one moves from the top
to bottom plots.}
\label{fig:photomodels}
\end{figure*}

The photoionization model results are presented in
Figure~\ref{fig:photomodels}, where we plot the ratio of the
model predictions to the observed emission-line fluxes.  With the
exception of cloud 1, the vast majority of the emission lines are
reproduced reasonably well (within a factor of two) by the models.
Given the simplicity of the models and the limited number of input
parameters, this agreement suggests that photoionization is the
dominant ionization mechanism in most of the NLR clouds.

High-excitation \NeVwa\ and \NeVwb\ lines were detected in the
spectra of the two NLR clouds closest to the nucleus (clouds 4 and
5).  The highly-ionized \NeV\ emission lines have a large ionization
potential (97.1~eV) and were not produced by our single-component,
optically-thick (radiation bounded) photoionization models.  The
presence of these high-ionization emission lines indicates that an
additional optically-thin (matter bounded) component is required in
these clouds, similar to that found by studies of other Seyferts
\citep{Binette1996, Komossa1997, Kraemer2000}.  The two-component
models were generated by assuming that all of the \NeV\ emission
arises in the high-ionization matter-bounded component and that lower
ionization lines (e.g. \OIII, \OII, \NII, and \SII) are produced in
the radiation-bounded component.  As before, these models were
optimized in Cloudy by allowing the ionization parameter, total
hydrogen density, and nitrogen abundance to vary.  For the
matter-bounded components, we also varied the total hydrogen column
density, which controls the depth of the cloud.  The resulting
two-component models for clouds 4 and 5 are shown in
Figure~\ref{fig:photomodels}.  Single-component, radiation-bounded
models were used for all of the other NLR clouds.  The model
parameters used to generate the photoionization models for each cloud
are presented in Table~\ref{tbl:cloudyparam}.

\begin{deluxetable*}{cccccc}
\tabletypesize{\footnotesize}
\tablecolumns{6}
\tablewidth{0pt}
\tablecaption{Photoionization Model Parameters \label{tbl:cloudyparam}}
\tablehead{\colhead{Cloud}
   &\colhead{$\log\,U$}
   &\colhead{Hydrogen Density}
   &\colhead{Column Density}
   &\colhead{Nitrogen Abundance}
   &\colhead{Note}\\
   \multicolumn{1}{c}{}
   & \multicolumn{1}{c}{}
   & \multicolumn{1}{c}{($10^{3}$ \ivol)}
   & \multicolumn{1}{c}{($10^{20}$ \iarea)}
   & \multicolumn{1}{c}{(N$_{\odot}$)}
   & \multicolumn{1}{c}{}
}
\startdata
                 1  &  $-$2.83  &  4.6  &  6.1  &  3.4  &  RB \\
                 2  &  $-$3.27  &  2.3  &  1.9  &  3.0  &  RB \\
                 3  &  $-$3.07  &  0.7  &  2.6  &  4.0  &  RB \\
4\tablenotemark{a}  &  $-$3.13  &  2.6  &  2.8  &  4.3  &  RB \\
                    &  $-$1.52  &  1.0  &  0.8  &  4.3  &  MB \\
5\tablenotemark{b}  &  $-$2.90  &  1.8  &  4.6  &  4.3  &  RB \\
                    &  $-$1.08  &  0.2  &  1.8  &  4.3  &  MB \\
                 6  &  $-$3.26  &  6.3  &  2.5  &  4.3  &  RB \\
                 7  &  $-$3.39  &  4.8  &  1.7  &  3.6  &  RB \\
                 8  &  $-$3.21  &  2.5  &  2.2  &  3.9  &  RB \\
                 9  &  $-$3.39  &  5.1  &  1.8  &  3.1  &  RB \\[-1em]
\enddata
\tablecomments{RB: radiation bounded; MB: matter bounded.}
\tablenotetext{a}{90\% RB + 10\% MB.}
\tablenotetext{b}{80\% RB + 20\% MB.}
\end{deluxetable*}

\begin{figure}[ht]
\epsscale{1.1}
\plotone{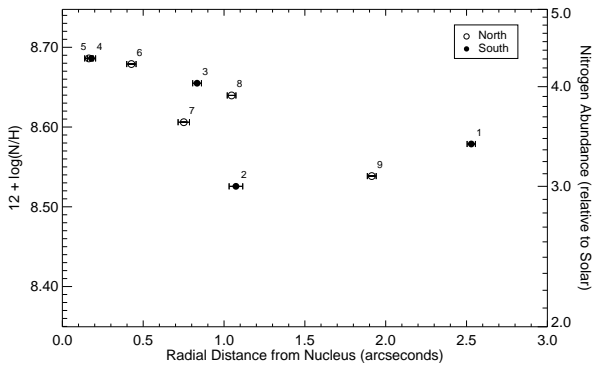}
\caption{Nitrogen abundance (relative to solar) of the optimized
Cloudy photoionization models for the NLR clouds.}
\label{fig:modelout}
\end{figure}

The model results show the presence of a gradient in both the
ionization parameter and the nitrogen abundance.  Excluding cloud 1,
which is poorly fit by the photoionization models, the ionization
parameter is larger for the clouds closest to the nucleus and
decreases with increasing radial distance out to cloud 9
($\sim$2\arcsec, 80~pc).  This result confirms the general trend
observed earlier in the \OIIIwb/\OIIw\ and \OIIIwb/\Hb\ emission-line
ratios and is consistent with the radial dilution of a central
ionizing radiation field in which the gas density decreases more
slowly than $r^{-2}$.  Interestingly, the photoionization model
results also show a gradient in nitrogen abundance over the NLR
region (Fig.~\ref{fig:modelout}).  The nitrogen abundance appears to
be larger closer to the nucleus ($\sim$4.5~N$_{\odot}$) and decreases
with increasing radius, reaching $\sim$3~N$_{\odot}$ for the outer
clouds.

\cite{Peimbert1968} was the first to claim that solar abundances
could not explain the observed emission-line ratios in the
near-nuclear region of M51.  His work suggested that nitrogen was
overabundant in the nucleus of M51 ($r\ \le\ 3\farcs5$) by a factor
of about five relative to solar.  Later emission-line studies by
\cite{Rose1982} and \citeauthor{Ford1985} (\citeyear{Ford1985},
$\sim$19~N$_{\odot}$ for the XNC) also inferred an enhancement of
nitrogen in the circumnuclear region of M51.  Our large
\NIIww/\Ha\ ratios (2.8 $-$ 6.2) confirm that nitrogen is enhanced in
the NLR of M51.

Nitrogen overabundances on the scale inferred by us in M51 have also
been claimed in studies of many Seyfert galaxies \citep{Storchi1989,
Storchi1990, Dopita1996, Storchi1998}, LINERs \citep{Binette1985,
Filippenko1992}, and starburst galaxies \citep{Coziol1999}.  In
particular, \citep{Storchi1990} found that a range of nitrogen
abundances, from 0.5 to 5 times solar, provided a better fit to the
the NLR emission-line ratios from a sample of 180 Seyfert~2s and
LINERs.  Of these galaxies, approximately one-third presented a
nitrogen overabundance in the nucleus \citep{Storchi1991}.  These
results suggest that nitrogen is overabundant in active nuclei
whenever \NIIww/\Ha\ $> 1.5$.  The enrichment of nitrogen found in
the nuclei of Seyferts may indicate that the N/O abundance gradient
found in the disks of spiral galaxies \citep{Searle1971, Smith1975,
Vila-Costas1993} continues into the nuclear regions.  Generally
speaking, the selective enhancement of nitrogen may be caused by
secondary CNO nucleosynthesis, in which nitrogen forms from
preexisting carbon and oxygen \citep{Thurston1996}, and may be
distributed by stellar winds from massive stars
\citep{CidFernandes1992}.

\subsection{Model Discrepancies \label{sec:modeldiscrep}}

While the overall agreement of the simple photoionization model
predictions and the observed optical spectrum generally is good,
there are a few discrepancies.  The poorest match occurs for the
\OIIw\ emission line, which is generally overpredicted by the models.
For three of the NLR clouds this discrepancy is within a factor of
2.5.  The largest mismatch (a factor of $\sim$3) occurs for cloud 6.
Although it is possible that our measured \OII\ fluxes are low as a
result of dereddening errors, additional dereddening would change
the emission-line fluxes for the entire spectrum.  In particular,
the \NeIIIww\ lines, which are at a comparable wavelength to \OIIw,
would also increase in strength counter to the model predictions.
Given the generally good agreement of the models with the dereddened
line fluxes, the discrepancy between the \OIIw\ model prediction
and the dereddened \OIIw\ flux does not appear to be a result of
dereddening errors.

One method to reduce the predicted \OIIw\ emission-line fluxes is to
increase the total hydrogen density, $n_H$.  \OIIw\ has a relatively
low critical density of $3.3 \times 10^{3}$~\ivol, above which the
lines are collisionally suppressed.  To examine this effect, we have
generated photoionization models with increased total hydrogen
density.  While the \OIIw\ lines are indeed suppressed, the higher
density models provide a poorer overall fit to the observed
spectrum.  In particular, \NI\ and \SII\ are suppressed and many
lines, including \NeIIIwa, \OIIIwb, \OIwa, and \SIIIwb, are
overpredicted for these models.  It is possible that our simple
assumption of constant density clouds is invalid and that more
complex models involving density stratification may provide a better
fit for \OIIw.  Or perhaps, the nitrogen overabundance described in
\S~\ref{sec:modelresults} is correlated with an oxygen
underabundance.

We also find modest discrepancies for the \SIII\ and \SII\ emission
lines.  \SIII\ is slightly overproduced (by a factor of $\sim$1.4)
and \SII\ is slightly underproduced (by a factor of $\sim$1.4) by the
models.  A general problem with photoionization models is that they
overpredict the strength of the \SIIIwa\ and \SIIIwb\ lines
\citep{Ho1993, Kraemer2000}.  \cite{Osterbrock1992} suggested that
this discrepancy may be improved with underabundances of sulfur,
however, we note that the dielectronic recombination rates for sulfur
are uncertain \citep{Ali1991}.  While the \SIII\ lines are
overpredicted, the \SIIwa\ and \SIIwb\ lines are a somewhat
underpredicted.  This in turn may suggest the presence of a
overabundance of sulfur, which has been detected by
\cite{Storchi1990} in a sample of 180 Seyfert and LINER galaxies.
Clearly, altering the sulfur abundances will not improve the fit for
both \SIII\ and \SII.  However, the discrepancy in these lines is
relatively modest.

As noted earlier, the photoionization models provide a poor match to
the observed spectrum of cloud 1.  This cloud also possesses the
largest \OIII/\Hb\ ratio, indicating the presence of high
ionization.  The proximity of this cloud to the entry point of the
radio jet into the lobe suggests that shock excitation may be an
important source of ionization.  Previous studies \citep{Ford1985,
Cecil1988, Terashima2001} have provided considerable evidence that
shocks contribute to the ionization structure of the XNC in M51.

\section{Shock Ionization Models}
\label{sec:shocks}

While it is widely held that the dominant source of energy input into
the NLR gas is photoionization by a nonstellar nuclear continuum
\citep{Koski1978, Osterbrock1989}, collisional shock wave heating may
provide an additional source of ionization in the NLR
\citep{Morse1996, Bicknell1998, Wilson1999}.  Shock waves generated
by supersonic radio ejecta would compress and heat the ambient gas
and may contribute, at least locally, a significant fraction of the
ionizing photons \citep{Dopita1995, Dopita1996, Bicknell1998}.  These
EUV photons are generated in the hot ($T \sim 10^{6}$~K) postshock
cooling region and diffuse both upstream and downstream.  The photons
diffusing downstream influence the ionization of the shock
recombination region, while those diffusing upstream ionize the
preshock medium.

\cite{Dopita1995, Dopita1996} have created a low-density grid of fast
radiative shock models using the code MAPPINGS~II.  The models assume
a steady-state one-dimensional radiative flow with roughly solar
abundances.  \cite{Dopita1995, Dopita1996} emphasized the importance
of the EUV photons that diffuse upstream into the preshock gas and
produce a high ionization precursor region.  We refer to the
MAPPINGS~II shock models that include the postshock cooling and
photoionized gas as ``shock-only'' models, while those that include
contributions from both the shocked gas and the ionized precursor
region as ``shock + precursor'' models.  Current, publicly available,
shock models are limited because they represent a one-dimensional
model of a more complex three-dimensional shock structure
\citep{Dopita1996}.

The two most important parameters in controlling the emergent shock
spectrum are the shock velocity, $v_s$, and the magnetic parameter,
\Bn.  The shock velocity controls the shape of the ionizing
continuum, and the magnetic parameter controls the effective
ionization parameter of the postshock recombination zone
\citep{Dopita1995, Dopita1996}.  The MAPPINGS~II model grids are
computed for shock velocities in the range of $v_s=150-500$~\kms\ and
magnetic parameters $\Bn=0-4$~\muGcm.  Magnetic parameters
$\Bn\sim2-4$~\muGcm\ correspond to the equipartition of the magnetic
and thermal pressures in the preshock medium \citep{Dopita1996}.

We have compared our observed NLR line ratios with the MAPPINGS~II
shock models to determine if shocks play a significant role in the
excitation of the NLR gas.  As demonstrated earlier with the Cloudy
photoionization models, there appears to be an overabundance of
nitrogen in the near-nuclear region of M51.  Similarly, we find that
the solar abundance MAPPINGS~II model grids cannot reproduce the
large observed \NII/\Ha\ ratios of the NLR clouds in M51.  For a
sample of active galaxies, \cite{Dopita1996} observed \NII/\Ha\
ratios that were $\sim$0.4~dex larger than the solar abundance
MAPPINGS~II shock model grids.  These authors likewise suggested that
this discrepancy was due to an enhancement of the nitrogen abundance
and generated new shock models with a N/O abundance of two times
solar.  They found a nearly one-to-one increase in the strength of
the \NII\ lines with the scaled abundance.  Thus to approximate the
effect of nitrogen enhancement in M51, we scaled the MAPPINGS~II
nitrogen emission-line fluxes (\NI\ and \NII) by the relative
nitrogen overabundance.

The MAPPINGS~II shock-only models are ruled out for all of our NLR
clouds because they fail to predict the observed \OIIw/\OIIIwb\ and
\OIIIwb/\Hb\ line ratios.  The M51 \OIIw/\OIIIwb\ line ratios lie in
the range 0.1$-$1.1, while the shock-only models predict
\OIIw/\OIIIwb\ $>$ 3.  This is not a surprising result as the
shock-only models generally lie in the region occupied by LINER
galaxies \citep{Dopita1995, Dopita1997}.

However, the MAPPINGS~II shock+precursor models can reproduce the
observed \linebreak \OIIw/\OIIIwb\ and \OIIIwb/\Hb\ line ratios for shock
velocities in the range $300-500$~\kms\ ($T_{X} \approx  1.3 - 3.5
\times 10^{6}$~K), consistent with results for several other Seyfert
galaxies \citep{Dopita1995, Allen1999, Ferruit1999a, Evans1999}.
Unfortunately, for shock velocities in this range there is
considerable overlap of the shock+precursor and photoionization model
grids for most of the optical emission lines.  As such, optical
emission lines generally are unable to provide conclusive
discrimination between shock and photoionization models
\citep[e.g,][]{Dopita1995, Allen1999, Evans1999}.  However, shocks
provide a rich, collisionally-excited UV spectrum, many of whose
lines are predicted to be much stronger than that found by
photoionization models \citep{Dopita1996, Allen1998}.  Because of
this, \cite{Allen1998} emphasized the importance of ultraviolet
emission lines, such as \ionw{N}{3}{991}, \CIVw, \ionw{He}{2}{1640},
\semiforbw{C}{3}{1909}, and \semiforbw{C}{2}{2326}, in distinguishing
between shock and photoionization models.  Unfortunately, the
ultraviolet emission in M51 is very weak, and a sensitive UV spectrum
of the near-nuclear region of M51 does not exist.  Furthermore, the
available {\em IUE} UV spectrum only shows a blended
\semiforbw{Si}{3}{1892} and \semiforbw{C}{3}{1909} emission line,
which has been attributed to \HII\ regions located within the large
\aslit{10}{20} aperture \citep{Ellis1982}.

\begin{figure}[ht]
\epsscale{1.2}
\plotone{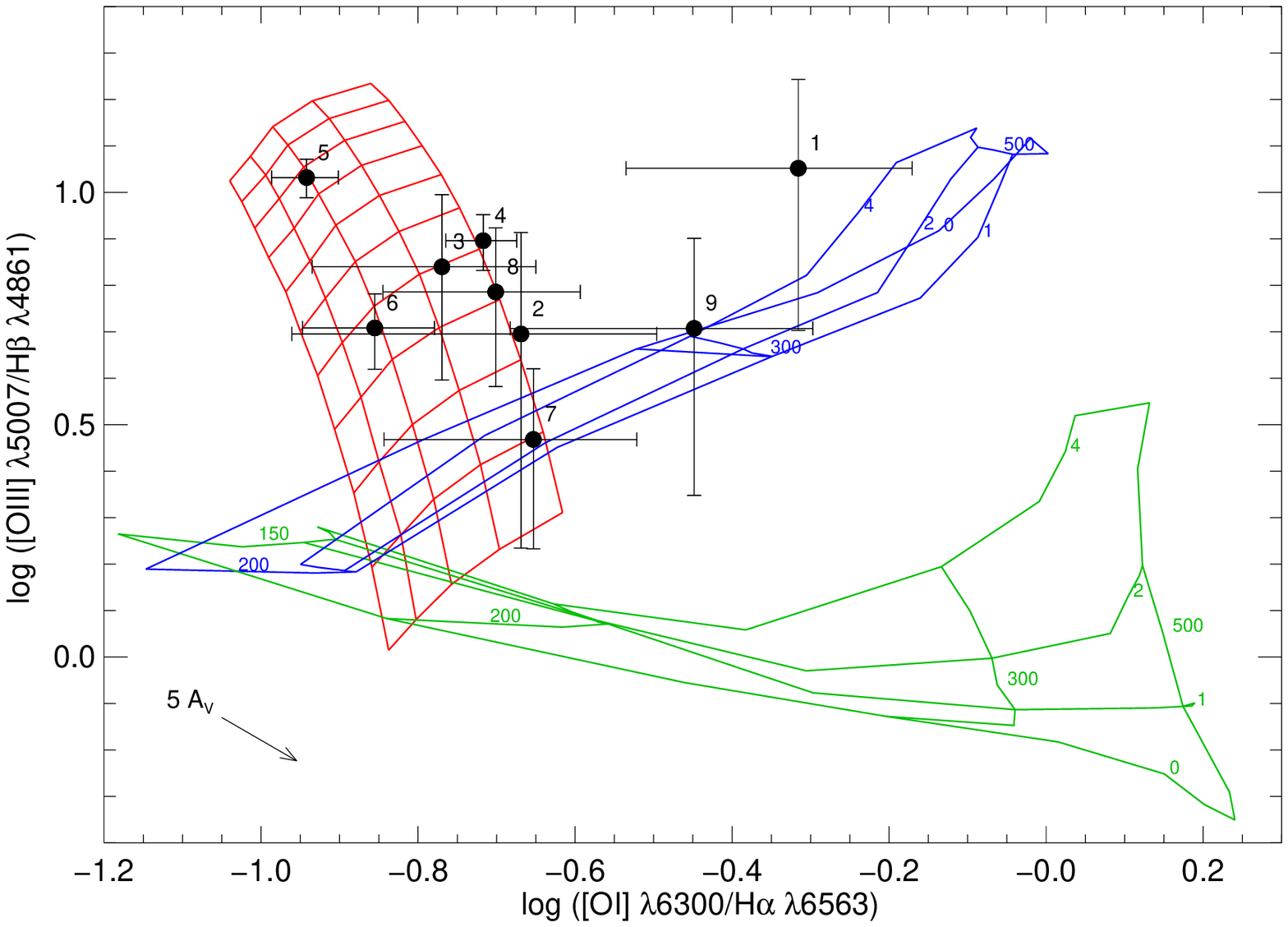}
\caption{\OIIIwb/\Hb\ vs. \OIwa/\Ha\ for the NLR clouds overplotted
with the MAPPINGS~II shock-only (green), shock+precursor (blue) and
Cloudy model grids with $4\times$ solar nitrogen (red).  The
MAPPINGS~II model grids are labeled with shock velocity in units of
\kms\ (shock-only:  150$-$500~\kms, shock+precursor: 200$-$500~\kms)
and magnetic parameter \Bn\ (0$-$4) in units of \muGcm.  The Cloudy
photoionization model grid spans $\log\ U$ in the range $-$3.5 to
$-$2.5 (bottom to top) and $\log\ n_{H}$ from 2 to 4~\ivol\ (left to
right).}
\label{fig:mapo1plot}
\end{figure}

Without UV emission lines, the best diagnostic we have available to
differentiate between the shock+precursor and photoionization models
is \OIIIwb/\Hb\ vs. \OIwa/\Ha\ (Fig.~\ref{fig:mapo1plot}).  As shown
in Fig.~\ref{fig:mapo1plot}, shock+precursor models with shock
velocities $v_s\ \ga\ 300$~\kms\ predict a stronger \OIwa/\Ha\ ratio
than our photoionization models.  Overall, our observed
\OIwa/\Ha\ line ratios appear to be more consistent with the
photoionization model grids.  Excluding clouds 1, 9, and 7, the
shock+precursor models overpredict the strength of the
\OIwa/\Ha\ ratio by factors of 2$-$6, effectively ruling out the
shock+precursor models for these clouds.

While the shock+precursor models are able to predict the strength of
the \OIwa/\Ha\ ratio in clouds 7 and 9, they fail to provide a good
fit for all of the observed emission lines.  For cloud 7, the best
overall fit shock+precursor model is for $v_s=250$~\kms\ and
$\Bn=1~\muGcm$.  However, in this model the strength of the high
ionization \HeIIw/\Hb\ ratio is underpredicted by a factor of five.
The \HeIIw/\Hb\ prediction is improved by increasing the shock
velocity to 350~\kms, but the \OIwa/\Hb\ and \NIw/\Hb\ ratios are
overproduced by a factor of $\sim$3 and the \OIIIwb/\Hb\ and
\OIIw/\Hb\ ratios are overproduced by factors $\ga$~2.  Turning to
cloud 9, the best overall shock+precursor model fit is also for
$v_s=250$~\kms\ and $\Bn=1~\muGcm$.  While this model provides the
lowest $\chi^2$, many of the emission lines, including \NeIIIwa,
\OIIIwb, and \NIIwb\ (even with 3.1~N$_{\odot}$), are underproduced
by factors of $\sim$2, and \HeIIw\ is underproduced by a factor of
four.  The fit appears to be influenced by the relatively strong
observed \OIIw, \NeIIIwa, \NIw, and \OIwa\ emission lines.  Excluding
these lines from the $\chi^2$ calculation, we find the best
shock+precursor model parameters are $v_s=350$~\kms\ and
$\Bn=1~\muGcm$.  While this model provides an improved fit for most
of the emission lines, \NIw\ is overproduced by a factor of four and
\OIIw, \OIwa, and \NeIIIwa\ are overproduced by $\sim$2.  There does
not appear to be a particular combination of $v_s$ and \Bn\ that
provides a good fit to all of the emission lines in cloud 9.  While
shocks cannot necessarily be ruled out in clouds 7 and 9 based on our
optical emission-line data, we find that our photoionization models
provide a better overall fit to these NLR clouds than the
shock+precursor models.

\subsection{Cloud 1}

\begin{figure}[ht]
\epsscale{1.2}
\plotone{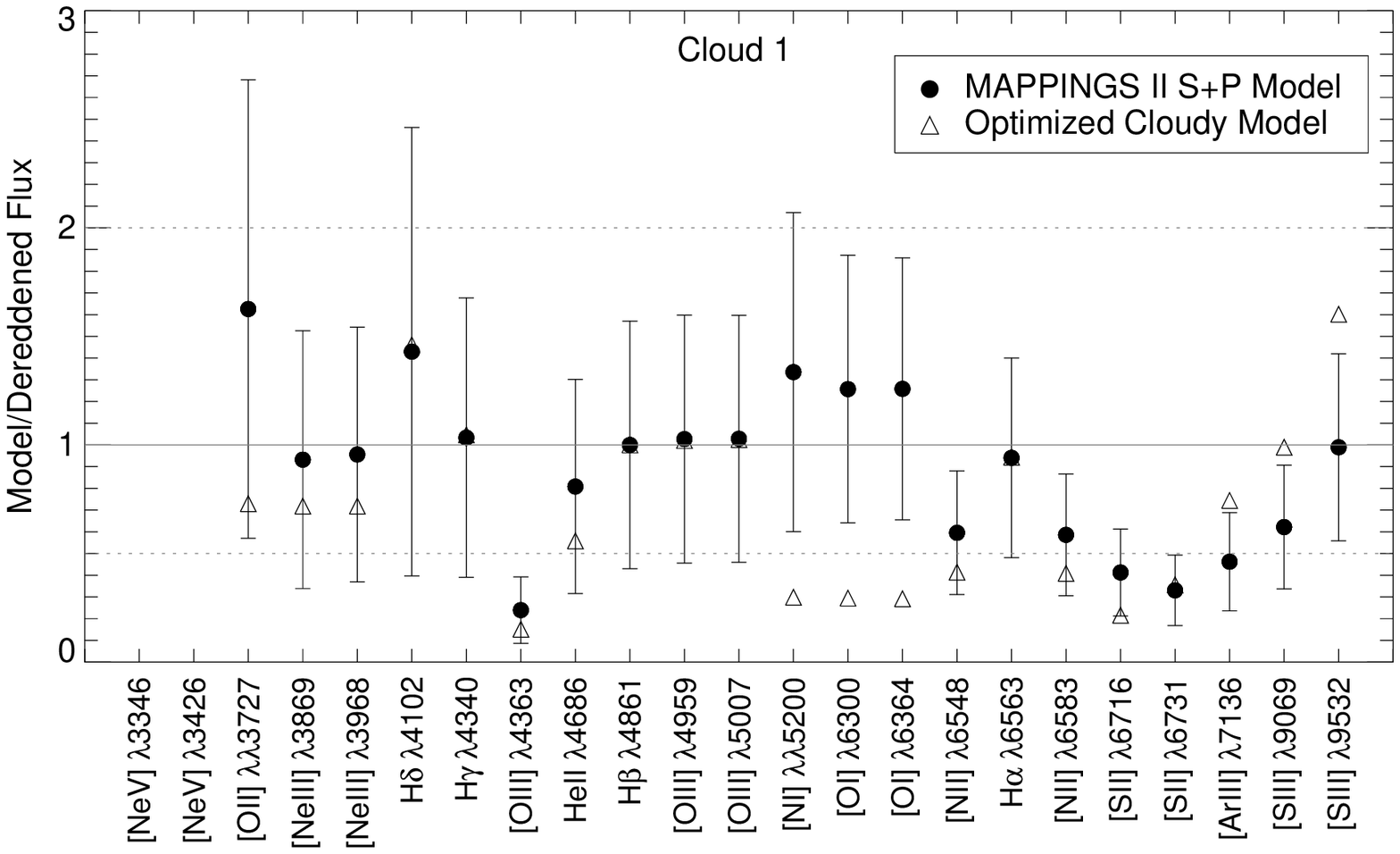}
\caption{Ratios of the MAPPINGS~II shock+precursor model predictions
and the optimized Cloudy photoionization model predictions to the
dereddened emission-line fluxes of cloud 1.  The MAPPINGS~II
shock+precursor model is for $v_s=450$~\kms\ and $\Bn=4~\muGcm$ with
the nitrogen lines scaled by 3.5 to approximate the effect of
nitrogen overabundance.}
\label{fig:cloud1best}
\end{figure}

In contrast, we find that shocks are important for cloud 1.  As
discussed in \S~\ref{sec:modeldiscrep}, photoionization models do not
provide a good fit to observed optical spectrum of cloud 1 (see
Fig~\ref{fig:photomodels}).  However, the MAPPINGS~II shock+precursor
models provide a reasonable fit to the observed emission lines in
this cloud.  The best fit shock+precursor models have parameters in
the range $v_s=400-450$~\kms\ and $\Bn=2-4~\muGcm$ with scaled
nitrogen abundances of $\sim$3.5.  In Fig.~\ref{fig:cloud1best} we
plot the ratio of the $v_s=450$~\kms\ and $\Bn=4~\muGcm$
shock+precursor model predictions to the observed emission-line
fluxes in cloud 1.  For comparison, we have also overplotted the
optimized Cloudy photoionization model results for this cloud.  In
particular, the shock+precursor models produce a better match to the
\NIw, \OIwa, and \OIwb\ lines than the photoionization models, which
underproduce these emission lines by factors of $\sim$3.  In
Table~\ref{tbl:cloud1map2}, we compare the optical spectrum of cloud
1 with the shock+precursor model predictions for the full range of
shock velocities and magnetic parameters.

\begin{deluxetable*}{lrccccc}
\tabletypesize{\footnotesize}
\tablecolumns{7}
\tablewidth{0pt}
\tablecaption{Line Ratios for Cloud 1 and MAPPINGS II
Shock+Precursor Models \label{tbl:cloud1map2}}
\tablehead{
   &  & \multicolumn{5}{c}{Shock+Precursor Models} \\
   \cline{3-7}\\[-0.8em]
   \colhead{Line\tablenotemark{a}}
   &\colhead{Cloud 1\tablenotemark{b}}
   &\colhead{200 \kms}
   &\colhead{300 \kms}
   &\colhead{400 \kms}
   &\colhead{450 \kms}
   &\colhead{500 \kms}
}
\startdata
\OIIw &  4.39 (2.85) & 2.29$-$3.64 & 1.69$-$5.19 & 1.73$-$6.97 & 1.70$-$7.14 & 1.70$-$7.19 \\
\NeIIIwa &  1.46 (0.93) & 0.18$-$0.21 & 0.37$-$0.58 & 0.71$-$1.13 & 0.93$-$1.36 & 1.16$-$1.59 \\
\NeIIIwb &  0.44 (0.27) & $\sim$0.06 & 0.11$-$0.18 & 0.22$-$0.35 & 0.29$-$0.42 & 0.36$-$0.49 \\
\Hdw &  0.18 (0.13) & $\sim$0.25 & 0.25$-$0.26 & $\sim$0.26 & $\sim$0.26 & $\sim$0.26 \\
\Hgw &  0.45 (0.28) & $\sim$0.46 & $\sim$0.46 & 0.46$-$0.47 & $\sim$0.47 & 0.46$-$0.47 \\
\OIIIwc\tablenotemark{c} &  0.04 (0.03) & 0.03$-$0.04 & $\sim$0.01 & $\sim$0.01 & $\sim$0.01 & $\sim$0.01 \\
\HeIIw &  0.41 (0.25) & $\sim$0.05 & 0.15$-$0.20 & 0.27$-$0.36 & 0.25$-$0.33 & 0.23$-$0.31 \\
\Hbw &  1.00 (0.57) & 1.00 & 1.00 & 1.00 & 1.00 & 1.00 \\
\OIIIwa &  3.92 (2.18) & 0.53$-$0.55 & 1.54$-$1.70 & 2.78$-$3.15 & 3.59$-$4.03 & 4.19$-$4.78 \\
\OIIIwb & 11.27 (6.23) & 1.53$-$1.58 & 4.44$-$4.90 & 8.01$-$9.09 & 10.33$-$11.59 & 12.08$-$13.76 \\
\NIw &  0.80 (0.44) & 0.17$-$0.27 & 0.56$-$0.80 & 0.87$-$1.33 & 1.07$-$1.50 & 1.39$-$1.64 \\
\OIwa &  1.49 (0.73) & 0.22$-$0.40 & 0.89$-$1.31 & 1.68$-$2.38 & 1.87$-$2.53 & 2.37$-$2.76 \\
\OIwb &  0.48 (0.23) & 0.07$-$0.13 & 0.29$-$0.42 & 0.54$-$0.77 & 0.60$-$0.82 & 0.76$-$0.89 \\
\NIIwa &  4.79 (2.29) & 0.68$-$1.46 & 0.82$-$2.26 & 1.17$-$2.86 & 1.28$-$2.95 & 1.34$-$2.94 \\
\Haw &  3.09 (1.51) & 2.99$-$3.02 & 2.92$-$2.95 & 2.90$-$2.91 & 2.90$-$2.91 & 2.89$-$2.92 \\
\NIIwb & 14.33 (6.85) & 2.01$-$4.31 & 2.40$-$6.66 & 3.44$-$8.41 & 3.77$-$8.69 & 3.93$-$8.66 \\
\SIIwa &  1.90 (0.92) & 0.29$-$0.56 & 0.58$-$0.91 & 0.81$-$1.21 & 0.78$-$1.16 & 0.82$-$1.11 \\
\SIIwb &  1.71 (0.84) & 0.25$-$0.41 & 0.49$-$0.70 & 0.58$-$1.14 & 0.57$-$1.23 & 0.59$-$1.27 \\
\ArIIIw &  0.49 (0.24) & 0.08$-$0.11 & 0.11$-$0.20 & 0.10$-$0.23 & 0.09$-$0.23 & 0.09$-$0.22 \\
\SIIIwa &  0.96 (0.44) & 0.24$-$0.33 & 0.25$-$0.49 & 0.24$-$0.56 & 0.24$-$0.60 & 0.25$-$0.65 \\
\SIIIwb &  1.47 (0.64) & 0.59$-$0.80 & 0.60$-$1.18 & 0.59$-$1.37 & 0.59$-$1.45 & 0.60$-$1.58 \\[-1em]
\enddata
\tablenotetext{a}{Normalized to \Hb\ unless otherwise indicated.}
\tablenotetext{b}{Dereddened fluxes with errors shown in parenthesis.}
\tablenotetext{c}{\OIIIwc/\OIIIwb\ electron temperature ratio.}
\tablecomments{The range of values in columns 3$-7$ correspond to the range
of magnetic parameters $\Bn = 0-4$~\muGcm.  The nitrogen lines have been
scaled by 3.5 to approximate the effect of an overabundance of nitrogen.}
\end{deluxetable*}

For a fully ionized gas with solar helium abundances, the predicted
postshock temperature can be estimated as $T_{s} = 1.4 \times
10^{5}\ v_{s100}^2$~K, where $v_{s100}$ is the shock velocity in
units of 100~\kms\ \citep{Hollenbach1979}.  For shock velocities in
the range $v_s=400-450$~\kms, we derive postshock temperatures in the
range from 2.2 $-$ 2.8 $\times 10^{6}$~K.  These values are roughly
consistent with the X-ray temperature of $T_{X} = 6.7 \times
10^{6}$~K obtained from \Chandra\ observations of the XNC
\citep{Terashima2001}, providing further evidence that shocks are
important for the excitation of cloud 1, which lies near the northern
edge of the XNC.
 
We do not find evidence of a broad \OIIIwb\ velocity dispersion for
cloud 1.  However, because this line is produced primarily in the
shock precursor region, this may be expected.  For lines which are
produced in the postshock region, such as \OIIw, \OIIIwc, and \NeVwb,
we only have lower spectral resolution data and are therefore unable
to determine if the velocity widths are broader for these lines.

The largest discrepancies between the MAPPINGS~II shock+precursor
model and the observed spectrum of cloud 1 are the underprediction of
the \SII\ and \OIIIwc\ emission lines.  As the \SIIIwa\ emission line
also is slightly underpredicted, it appears the model fit could be
improved with a modest overabundance ($\sim2\times$ solar) of
sulfur.  An overabundance of sulfur previously has been found in a
sample of Seyferts and LINERs and may be correlated with the nitrogen
enhancement \citep{Storchi1989, Storchi1990, Schmitt1994}.  However,
the discrepancy with the \SII\ lines may be a result of the
incomplete dielectronic recombination rates for sulfur
\citep{Ali1991}.

While the \OIIIwc/\OIIIwb\ ratio is underpredicted in cloud 1, this
appears to be a general feature of the MAPPINGS~II shock+precursor
models \citep{Dopita1995}.  This is especially true for shock
velocities $v_s\ \ga\ 300$~\kms, in which the majority of the
\OIII\ emission is produced in the lower-temperature photoionized
precursor rather than the hot postshock region \citep{Sutherland1993,
Dopita1995, Dopita1996}.  The \OIIIwc/\OIIIwb\ prediction is improved
with shock-only or low-velocity shock+precursor models, however,
these models provide a much worse overall fit to the ensemble of
emission lines (see Table~\ref{tbl:cloud1map2}).  In any case, the
MAPPINGS~II shock+precursor models match the observed optical
spectrum of cloud 1 better than the photoionization models (see
Fig.~\ref{fig:cloud1best}).

\subsection{\mbox{\Hb} Luminosity}

Correcting for the distance to M51 (8.4~Mpc), the observed cloud 1
\Hb\ luminosity is $2.7\ \times\ 10^{36}$~\ergs.  The \Hb\ luminosity
predicted from the shock+precursor models is given by
\cite{Dopita1995} as
\begin{eqnarray}
\label{eqn:hbshock}
L(\Hb) \ =\ A\ n_0 & \left[7.44\times10^{-6}\,
\left(\frac{\mbox{$v_s$}}{\mbox{100\ km$\,$s$^{-1}$}}\right)^{2.41} \ + \ \right. \quad\quad\quad \nonumber \\
 & \quad\quad \left. 9.85\times10^{-6}\, \left(\frac{\mbox{$v_s$}}{\mbox{100\ km$\,$s$^{-1}$}}\right)^{2.28} \right] \ ,
\end{eqnarray}

\noindent
where $A$ is the shock area (\iarea) and $n_0$ (cm$^{-3}$) is the
preshock density.  Because the cloud 1 shock is viewed nearly edge on
(see \S~\ref{sec:ioncone}), we follow the examples of
\cite{Ferruit1999b} and \cite{Schmitt2002} and estimate the shock
area by assuming a line-of-sight depth equal to the slit width
(0\farcs19, 8~pc).  For this cylindrical symmetry we derive a shock
area of $4.5 \times1 0^{38}$~\iarea\ (47~pc$^2$).  We estimate the
preshock density using the shock compression factors for the \SII\
zone computed by \cite{Ferruit1999b}.  For the best-fitting shock
velocities in the range from $400-450$~\kms\ these compression
factors are $46.4-50$.  Given the calculated \SII\ electron density
of 490~\ivol\ for cloud 1, the resulting preshock densities are in
the range from $9.8 - 10.6$~\ivol.  For shock velocities
$v_s=400-450$~\kms, the shock+precursor model predicts \Hb\
luminosities in the range $L(\Hb)=2.1-2.6 \times 10^{36}$~\ergs,
which is approximately 78$-$96\% of the observed \Hb\ luminosity of
cloud 1.  This suggests that, within the uncertainties in the
preshock density and the estimated shock velocity, shocks are the
dominant excitation mechanism in cloud 1.

\section{Summary}
\label{sec:summary}

We have examined the physical conditions in the NLR of M51 using
archival \HST/WFPC2 narrowband images, high spatial resolution,
long-slit \HST/STIS spectra, and VLA 8.4~GHz radio continuum maps.
Our velocity measurements of the NLR clouds show that the radial
velocity distribution is inconsistent with gravitational rotation in
a circumnuclear disk.  The kinematics are consistent with radial
outflows from the nucleus that lie nearly in the plane of the sky.
On the southern side of the nucleus, the radial velocities of the NLR
clouds appear to be influenced by the presence of the southern radio
jet.

Through the use of standard line-ratio diagnostics, we have shown
that the emission-line ratios in the NLR clouds are in agreement with
those typically found in Seyfert galaxies.  The ionization state of
the gas, as probed by the \OIII/\Hb\ and \OIII/\OII\ line ratios,
decreases radially from the nucleus out to $r \sim 2\arcsec$.  This
is consistent with photoionization from a nonstellar nuclear
continuum source in which the density falls off more gradually than
$r^{-2}$.  The ionization is larger for cloud 1 ($r = 2\farcs5$) and
appears to be caused by the presence of shocks.

The \SII\ $\lambda6716/\lambda6731$ flux ratios indicate that the
average electron density within the NLR gas is $\sim$770~\ivol.
Given the large error bars for the \SII\ fluxes, no radial trend can
be inferred for the electron density.  Cloud 1 is the only cloud in
which we detected \OIIIwc.  The derived \OIII\ electron temperature
is 24,000~K.  This, coupled with our radio observations, suggests
that shocks are an important heating mechanism for this cloud.  Upper
limits to the \OIII\ electron temperature for clouds 4, 5, 6, and 8
($T_e\ \la\ 11,000$~K) indicate that collisional heating is less
important for these clouds.  These electron temperatures are
consistent with photoionization.

The optical and 3.6~cm radio emission both exhibit an extended,
knotty emission structure along a position angle of 170\dg.  Our
3.6~cm data are in agreement with the earlier 6~cm observations of
\cite{Crane1992}.  Both observations exhibit a weak southern jet,
$\sim$2\farcs5 in extent, that connects with a diffuse lobe structure
south of the nucleus.  The well-known XNC optical structure south of
the nucleus appears to be coincident with the southern radio lobe.
The higher spatial resolution 3.6~cm data exhibit several knots of
emission, indicating that there is structure in the jet.  In general,
these radio knots lie adjacent to the observed optical clouds,
suggesting that the radio jet may contribute to the kinematics or
ionization structure of the optical emission.  Cloud 1, at the
northern edge of the XNC, appears to lie near, but west of, the entry
point of the jet into the southern radio lobe.  Recent {\em Chandra}
observations \citep{Terashima2001} also detect X-ray emission
coincident with the XNC.

At the assumed entry point of the radio jet into the diffuse lobe
structure, we find both radio and optical emission.  The optical
emission, cloud 1, is straddled by two radio knots, R0 and R1.  The
proximity of cloud 1 with the jet-lobe entry point and radio knots R0
and R1 suggests that collisional excitation may be important for this
cloud.

Cloud-by-cloud photoionization models were generated to determine the
relative importance of photoionization for each cloud.  With the
exception of cloud 1, the photoionization models provide good fits to
the NLR emission-line clouds.  However, our models require supersolar
nitrogen abundances to match the dereddened \NIw, \NIIwa, and
\NIIwb\ line ratios.  This result is supported by other studies of
Seyfert galaxies that also required selective enhancement of nitrogen
by factors up to five times solar to fit the observed nitrogen
emission.  In addition, we find a strong gradient in the nitrogen
abundance in the near-nuclear region.  The nitrogen abundance is
$\sim$4.5~N$_{\odot}$ for the near-nuclear clouds ($r \sim 0\farcs2$)
and decreases for increasing radial distance to $\sim$3~N$_{\odot}$
for the outer clouds ($r \sim 2\arcsec$).

We also find a gradient in ionization parameter, which agrees with
the result inferred from the observed emission-line ratios.  We
detect the highly-ionized \NeVwa\ and \NeVwb\ emission lines in
clouds 4 and 5 ($r \sim 0\farcs2$), which indicates the presence of a
high-ionization matter-bounded component.  We successfully modelled
these clouds with two-component models, which included contributions
from both matter-bounded and radiation-bounded components.

The photoionization models do not provide a good fit to the observed
optical spectrum of cloud 1.  Given the prior evidence that shocks
contribute to the ionization structure of the XNC \citep{Ford1985,
Cecil1988}, we compared the emission-line fluxes of cloud 1 with
MAPPINGS~II shock model grids \citep{Dopita1995, Dopita1996}.  The
MAPPINGS~II shock+precursor models with velocities in the range
$v_s=400-450$~\kms, magnetic parameters $\Bn=2-4~\muGcm$, and scaled
nitrogen abundances of $\sim$3.5 provide a reasonable fit to the
observed emission lines in this cloud.  Later {\em Chandra} X-ray
observations by \cite{Terashima2001} further support the conclusion
that shocks are the dominant source of ionization at the location of
cloud 1.

\section{Conclusions}
\label{sec:conclusions}

The optical and radio morphologies in the NLR of M51 exhibit a
striking correspondence.  The conical, limb-brightened optical
morphology south of the nucleus is suggestive of evacuation by the
weak southern radio jet.  The near coincidence of the bright optical
emission clouds with the position angle of the radio jet and the
proximity of the optical and radio knots also suggest that the radio
jet may influence the kinematic and ionization structure of the
optical emission.

We find that, in general, photoionization is the dominant excitation
mechanism for the brightest optical emission-line clouds.  However,
shock+precursor models provide the best fit to the optical
emission-line cloud that lies near, but west of, the entry point of
the jet into the diffuse radio lobe.  Thus, shocks are an important
excitation mechanism for at least some regions in the NLR of M51.
Furthermore, we find some evidence for the presence of weak, higher
velocity dispersion clouds.  Several \OIII\ clouds were not well fit
by a single Gaussian and required a second, broader Gaussian emission
component to obtain a good fit.  The signal-to-noise and general
spectral resolution of our data were insufficient to permit a more
detailed analysis of this diffuse emission.  Higher signal-to-noise,
higher spectral resolution STIS data (Ferruit et al., in preparation)
that map the optical emission with four slit positions will help
determine if shocks play a more pervasive role in the NLR of this
Seyfert galaxy.  Additionally, higher signal-to-noise and higher
spatial resolution radio maps would be beneficial in determining the
importance of weak radio jet structures, observed in several Seyfert
galaxies, to the general AGN paradigm.

\acknowledgments
Support for these observations was provided by NASA contract
NAS5-30403 to LDB and MEK.  Support for WAB was provided by
Westerbork Observatory, which is operated by ASTRON (Netherlands
Foundation for Research in Astronomy) with support from the
Netherlands Foundation for Scientific Research (NWO).  This research
has made use of NASA's Astrophysics Data System Bibliographic
Services.  This research has also made use of the NASA/IPAC
Extragalactic Database (NED), which is operated by the Jet Propulsion
Laboratory, California Institute of Technology, under contract with
the National Aeronautics and Space Administration.


\end{document}